\documentclass{ws-mpla}

\usepackage{amsmath}
\usepackage{amssymb}

\newcommand{\jpsi}{J/$\psi$}
\newcommand{\psip}{$\psi^\prime$}
\newcommand{\chiAll}{$\chi$}

\newcommand{\chiOne}{$\chi_{1}$}
\newcommand{\chiTwo}{$\chi_{2}$}
\newcommand{\chicAll}{$\chi_c$}

\newcommand{\upsAll}{$\Upsilon$}
\newcommand{\upsOneS}{$\Upsilon(1S)$}
\newcommand{\upsTwoS}{$\Upsilon(2S)$}
\newcommand{\upsThreeS}{$\Upsilon(3S)$}
\newcommand{\chibAll}{$\chi_b$}

\newcommand{\pt}{$p_{\rm T}$}

\newcommand{\ccbar}{$c \bar{c}$}
\newcommand{\bbbar}{$b \bar{b}$}

\begin{document}

\markboth{Pietro Faccioli} {Questions and prospects in quarkonium polarization
measurements from pp to AA collisions}

\catchline{}{}{}{}{}

\title{QUESTIONS AND PROSPECTS IN QUARKONIUM POLARIZATION MEASUREMENTS FROM PROTON-PROTON
TO NUCLEUS-NUCLEUS COLLISIONS\footnote{Invited brief review, reflecting
seminars and presentations made at CERN, Fermilab, Brookhaven, DESY, HEPHY,
etc. Published in Modern Physics Letters A Vol. 27, No. 23 (2012) 1230022, doi:
10.1142/S0217732312300224, copyright World Scientific Publishing Company,
http://www.worldscinet.com/mpla/mpla.shtml}}

\author{\footnotesize PIETRO FACCIOLI}

\address{%
Laborat\'orio de Instrumenta\c{c}\~ao e F\'{\i}sica
Experimental de Part\'{\i}culas, 1000-149 Lisbon, Portugal\\
$^2$Centro de F\'{\i}sica Te\'orica de Part\'{\i}culas, 1049-001 Lisbon, Portugal\\
$^3$Physics Department, Instituto Superior T\'ecnico, 1049-001 Lisbon, Portugal\\
Pietro.Faccioli@cern.ch}

\maketitle

\pub{Received (Day Month Year)}{Revised (Day Month Year)}

\begin{abstract}

Polarization measurements are the best instrument to understand how
quark and antiquark combine into the different quarkonium states,
but no model has so far succeeded in explaining the measured \jpsi\
and \upsAll\ polarizations. On the other hand, the experimental data
in proton-antiproton and proton-nucleus collisions are inconsistent,
incomplete and ambiguous. New analyses will have to properly address
often underestimated issues: the existence of azimuthal
anisotropies, the dependence on the reference frame, the influence
of the experimental acceptance on the comparison with other
measurements and with theory. Additionally, a recently developed
frame-invariant formalism will provide an alternative and often more
immediate physical viewpoint and, at the same time, will help
probing systematic effects due to experimental biases. The role of
feed-down decays from heavier states, a crucial missing piece in the
current experimental knowledge, will have to be investigated.
Ultimately, quarkonium polarization measurements will also offer new
possibilities in the study of the properties of the quark-gluon
plasma.

\keywords{Quarkonium; polarization; QCD.}
\end{abstract}

\ccode{PACS Nos.: 11.80.Cr, 12.38.Qk, 13.20.Gd, 13.85.Qk}

\section{The experimental situation}
\label{sec:exp}

Quarkonia, bound states made of a quark (of type charm or beauty) and its
antiquark, offer us a privileged window over the physics of the strong force,
which is at the origin of visible matter and, yet, is the least well-understood
aspect of the Standard Model of elementary interactions. Quarkonia represent
the most elementary manifestation of the strong binding force and allow us to
study crucial open questions: how are quarks confined inside hadrons? How do
strong forces generate the properties of particles made of quarks? Can quarks
become unbound under extreme conditions (high temperature and density: the
quark-gluon plasma), as they existed in the first moments of the universe? To
test and consolidate the current theory of the strong force, quantum
chromo-dynamics (QCD), it is crucial to study how quarkonia are produced in
elementary (proton-proton) collisions and in the much more complex
nucleus-nucleus collisions, where the potential that binds the quarks and the
gluons should be screened and the medium should reflect the partonic degrees of
freedom.
However, our present understanding of this physics topic is rather
limited, despite the multitude of experimental data accumulated over
more than 30 years.\cite{bib:YellowRep-QWG}
The \jpsi\ and \psip\ direct production cross sections measured (in the mid
1990's) by CDF, in ${\rm p}\bar{\rm p}$ collisions at
1.8~TeV,\cite{bib:cdf1-psis} were seen to be around 50 times larger than the
available expectations, based on leading order calculations made in the scope
of the Colour Singlet Model (CSM). The non-relativistic QCD (NRQCD)
framework,\cite{bib:NRQCD} where quarkonia can also be produced as
\emph{coloured} quark pairs, succeeded in describing the measurements, opening
a new chapter in the studies of quarkonium production physics. However, these
calculations depend on non-perturbative parameters, the long distance colour
octet matrix elements, which have been freely adjusted to the data, thereby
decreasing the impact of the resulting agreement between data and calculations.
More recently, calculations of next-to-leading-order (NLO) QCD
corrections to colour-singlet quarkonium production showed an
important increase of the high-\pt\ rate, significantly decreasing
the colour-octet component needed to reproduce the quarkonium
production cross sections measured at the
Tevatron.\cite{bib:lansberg-HP08}
Given this situation, differential cross sections are clearly
insufficient information to ensure further progress in our
understanding of quarkonium production. Polarization measurements,
determining the average angular momentum states of the produced
quarkonia from their decay distributions, can provide the definitive
tests of the theory of quarkonium production. No other study
addresses more directly the question: how does the observed
quark-antiquark bound states acquire their final quantum numbers? In
fact, the competing mechanisms dominating in the different
theoretical approaches lead to very different expected polarizations
of the produced quarkonia at high \pt. NRQCD
predicts\cite{bib:BK,bib:Lei,bib:BKL} almost fully \emph{transverse}
polarization (angular momentum component $J_z = \pm 1$) for directly
produced \psip\ and \jpsi\ mesons with respect to their own momentum
direction (the \emph{helicity frame}), while according to the new
NLO calculations of colour-singlet quarkonium
production\cite{bib:lansberg-HP08} these states should show a strong
\emph{longitudinal} ($J_z = 0$) polarization component.

Having two very different theoretical predictions appears to be an ideal
situation in the prospect of discriminating between the two theory frameworks
using experimental data. However, the present experimental knowledge is
incomplete and contradictory.
A significant fraction (around one third\cite{bib:feeddown}) of
promptly produced \jpsi\ mesons (i.e.\ excluding contributions from
B hadron decays) comes from \chicAll\ and \psip\ feed-down decays.
This sizeable source of indirectly produced \jpsi\ mesons is not
subtracted from the current measurements, and its kinematic
dependence is not precisely known.
Despite this limitation, it seems safe to say that the pattern
measured by CDF\cite{bib:CDFpol2} of a slightly longitudinal
polarization of the inclusive prompt \jpsi\ is incompatible with any
of the two theory approaches mentioned above.
The situation is further complicated by the intriguing lack of
continuity between fixed-target and collider results, which can only
be interpreted in the framework of some specific (and speculative)
assumptions still to be tested.\cite{bib:pol}

The \bbbar\ system should satisfy the non-relativistic approximation much
better than the \ccbar\ case. For this reason, the \upsAll\ data are expected
to represent the most decisive test of NRQCD. However, the present data from
Tevatron,\cite{bib:upsCDF1,bib:upsCDF2,bib:upsD0} for $\langle p_{\rm T}
\rangle \le 27$~GeV$/c$, tend to contradict the crucial NRQCD hypothesis that
high-\pt\ quarkonia, produced by the fragmentation of an outgoing (almost
on-shell) gluon, are fully transversely polarized along their own direction.
At lower energy and \pt, the E866 experiment\cite{bib:e866_Ups} has
shown yet a different polarization pattern: the \upsTwoS\ and
\upsThreeS\ states have \emph{maximal} transverse polarization, with
no significant dependence on transverse or longitudinal momentum,
\emph{with respect to the direction of motion of the colliding
  hadrons} (Collins--Soper frame\cite{bib:CollinsSoper}). Unexpectedly, the \upsOneS, whose
spin and angular momentum properties are identical to the ones of
the heavier \upsAll\ states, is, instead, found to be only weakly
polarized. These results give interesting physical indications.
First, the maximal polarization of \upsTwoS\ and \upsThreeS\ along
the direction of the interacting particles places strong constraints
on the topology and spin properties of the underlying elementary
production process. Second, the small \upsOneS\ polarization
suggests that the bottomonium family may have a peculiar feed-down
hierarchy, with a very significant fraction of the lower mass state
being produced indirectly; at the same time, the polarization of the
\upsAll's coming from \chibAll\ decays should be substantially
different from the polarization of the directly produced ones.

This rather confusing situation demands a significant improvement in
the accuracy and detail of the polarization measurements, ideally
distinguishing between the properties of directly and indirectly
produced states.
We remind that the lack of a consistent description of the polarization
properties represents today's biggest uncertainty in the simulation of the LHC
quarkonium production measurements and represents the largest contribution to
the systematic error affecting the measurements of quarkonium production cross
sections and kinematic distributions.

It is true that measurements of the quarkonium decay angular distributions are
challenging, multi-dimensional kinematic problems, requiring large event
samples and a very high level of accuracy in the subtraction of spurious
kinematic correlations induced by the detector acceptance. The complexity of
the experimental problems which have to be faced in the polarization
measurements is testified, for example, by the disagreement between the CDF
\jpsi\ results obtained in Run~I\cite{bib:CDFpol1} and Run~II\cite{bib:CDFpol2}
and by the contradictory \upsOneS\ results obtained by CDF\cite{bib:upsCDF1}
and D0\cite{bib:upsD0}.
However, it is also true, as we shall emphasize hereafter, that most
experiments have presented in the published reports only a fraction of the
physical information derivable from the data. This happens, for example, when
the measurement is performed in only one polarization frame and is limited to
the polar projection of the decay angular distribution. As we have already
argued in Ref.~\refcite{bib:pol}, these incomplete measurements do not allow
definite physical conclusions.  At best, they confine such conclusions to a
genuinely model-dependent framework. Moreover, such a fragmentary description
of the observed physical process obviously reduces the chances of detecting
possible biases induced by not fully controlled systematic effects.

In this work we focus our attention on aspects that need to be taken in
consideration in the analysis of the data, so as to maximize the physical
significance of the measurement and provide all elements for its unambiguous
interpretation within any theoretical framework
(Sects.~\ref{sec:def}--\ref{sec:feeddown}).

We also discuss (Sect.~\ref{sec:seqsuppr}) how the polarization of vector
quarkonia, measured from dilepton event samples, can be used as an instrument
to study the suppression of $\chi_c$ and $\chi_b$ in heavy-ion collisions,
where a direct determination of signal yields involving the identification of
low-energy photons is essentially impossible.

\section{Basic concepts}
\label{sec:def}

Because of angular momentum conservation and basic symmetries of the
electromagnetic and strong interactions, a particle produced in a
certain superposition of elementary mechanisms may be observed
preferentially in a state belonging to a definite subset of the
possible eigenstates of the angular momentum component $J_z$ along a
characteristic quantization axis. When this happens, the particle is
said to be polarized. In the dilepton decay of quarkonium, the
geometrical shape of the angular distribution of the two decay
products (emitted back-to-back in the quarkonium rest frame)
reflects the average polarization of the quarkonium state. A
spherically symmetric distribution would mean that the quarkonium
would be, on average, unpolarized. Anisotropic distributions signal
polarized production.

The measurement of the distribution requires the choice of a
coordinate system, with respect to which the momentum of one of the
two decay products is expressed in spherical coordinates.  In
inclusive quarkonium measurements, the axes of the coordinate system
are fixed with respect to the physical reference provided by the
directions of the two colliding beams as seen from the quarkonium
rest frame.  The polar and azimuthal angles $\vartheta$ and
$\varphi$ describe the direction of one of the two decay products
(e.g.\ the positive lepton) with respect to the chosen polar axis
and to the plane containing the momenta of the colliding beams
(``production plane''). The actual definition of the decay reference
frame with respect to the beam directions is not unique.
Measurements of the quarkonium decay distributions used mainly two
different conventions for the orientation of the polar axis: the
flight direction of the quarkonium itself in the centre-of-mass of
the colliding beams (centre-of-mass helicity frame, HX) and the
bisector of the angle between one beam and the opposite of the other
beam (Collins--Soper frame, CS). The motivation of the latter
definition is that, in hadronic collisions, it coincides with the
direction of the relative motion of the colliding partons, when
their primordial transverse momenta, $k_{\rm T}$, are neglected.
We note that these two frames differ by a rotation of $90^\circ$
around the $y$ axis when the quarkonium is produced at high \pt\ and
negligible longitudinal momentum ($p_{\rm T} \gg |p_{\rm
  L}|$). All definitions become coincident in the limit of zero
quarkonium \pt. In this limit, moreover, for symmetry reasons any
azimuthal dependence of the decay distribution is physically
forbidden.

The most general decay angular distribution for inclusively observed
quarkonium states can be written as\cite{bib:qqbarExpClar}
\begin{eqnarray}
  W(\cos \vartheta, \varphi) \, \propto \, \frac{1}{(3 + \lambda_{\vartheta})} \,
  (1 + \lambda_{\vartheta} \cos^2 \vartheta
  +  \lambda_{\varphi} \sin^2 \vartheta \cos 2 \varphi +
  \lambda_{\vartheta \varphi} \sin 2 \vartheta \cos \varphi ) \, , \label{eq:observable_ang_distr}
\end{eqnarray}
where the three parameters $\lambda_{\vartheta}$ (``polarization''),
$\lambda_{\varphi}$ and $\lambda_{\vartheta \varphi}$ satisfy the
relations\cite{bib:triangles}
\begin{eqnarray}
&&|\lambda_\varphi| \le \frac{1}{2}\, (1 + \lambda_\vartheta ) \, ,
\quad
\lambda_\vartheta^2 + 2 \lambda_{\vartheta \varphi}^2 \le 1 \, , \nonumber \\
&& |\lambda_{\vartheta \varphi}| \le \frac{1}{2}\, (1 - \lambda_\varphi ) \, , \label{eq:triangles}\\
&& (1 + 2 \lambda_\varphi)^2 + 2\lambda_{\vartheta \varphi}^2 \le 1
\;\;\; \mathrm{for} \;\;\; \lambda_\varphi < -1/3 \,  , \nonumber
\end{eqnarray}
which, in particular, imply $|\lambda_\varphi| \le 1 $,
$|\lambda_{\vartheta \varphi}| \le \sqrt{2}/2$, $|\lambda_\varphi|
\le 0.5$ for $\lambda_{\vartheta} = 0$ and $\lambda_\varphi \to 0$
for $\lambda_{\vartheta} \to -1$.

\section{The importance of the reference frame and of the azimuthal anisotropy}
\label{sec:refframe}

The coefficients $\lambda_{\vartheta}$, $\lambda_{\varphi}$ and
$\lambda_{\vartheta \varphi}$ depend on the polarization frame. To illustrate
the importance of the choice of the polarization frame, we consider specific
examples assuming, for simplicity, that the observation axis is perpendicular
to the natural axis. This case is of physical relevance since when the decaying
particle is produced with small longitudinal momentum ($|p_{\rm L}| \ll p_{\rm
T}$, a frequent kinematic configuration in collider experiments) the CS and HX
frames are actually perpendicular to one another. In this situation, a natural
``transverse'' polarization ($\lambda_\vartheta = +1$ and $\lambda_\varphi =
\lambda_{\vartheta\varphi} = 0$), for example, transforms into an observed
polarization of opposite sign (but not fully ``longitudinal''),
$\lambda^\prime_\vartheta = -1/3$, with a significant azimuthal anisotropy,
$\lambda^\prime_\varphi = 1/3$. In terms of angular momentum wave functions, a
state which is fully ``transverse'' with respect to one quantization axis ($|J,
J_z \rangle = |1, \pm 1 \rangle$) is a coherent superposition of 50\%
``transverse'' and 50\% ``longitudinal'' components with respect to an axis
rotated by $90^\circ$:
\begin{equation}
|1, \pm 1 \rangle \quad \xrightarrow{ 90^\circ } \quad \frac{1}{2}
\; |1, +1 \rangle \; + \; \frac{1}{2} \; |1, -1 \rangle \; \mp \;
\frac{1}{\sqrt{2}} \; |1, 0 \rangle  \, .
\end{equation}
The decay distribution of such a ``mixed'' state is azimuthally
anisotropic. The same polar anisotropy $\lambda^\prime_\vartheta =
-1/3$ would be measured in the presence of a mixture of \emph{at
least two different production processes} resulting in 50\%
``transverse'' ($|J, J_z \rangle = |1, \pm 1 \rangle$) and 50\%
``longitudinal'' ($|J, J_z \rangle = |1, 0 \rangle$) natural
polarization along the chosen axis. In this case, however, no
azimuthal anisotropy would be observed.
As a second example, we note that a fully ``longitudinal'' natural
polarization ($\lambda_\vartheta = -1$) translates, in a frame
rotated by $90^\circ$ with respect to the natural one, into a fully
``transverse'' polarization ($\lambda^\prime_\vartheta = +1$),
accompanied by a maximal azimuthal anisotropy
($\lambda^\prime_\varphi = -1$). In terms of angular momentum, the
measurement in the rotated frame is performed on a coherent
admixture of states,
\begin{equation}
|1, 0 \rangle \quad \xrightarrow{\; 90^\circ \;} \quad
\frac{1}{\sqrt{2}} \; |1, +1 \rangle  - \frac{1}{\sqrt{2}} \; |1, -1
\rangle  \, ,
\end{equation}
while a \emph{natural} ``transverse'' polarization would originate
from the statistical superposition of \emph{uncorrelated} $|1, +1
\rangle$ and $|1, -1 \rangle$ states. The two physically very
different cases of a natural transverse polarization observed in the
natural frame and a natural longitudinal polarization observed in a
rotated frame are experimentally indistinguishable when the
azimuthal anisotropy parameter is integrated out. These examples
show that a measurement (or theoretical calculation) consisting only
in the determination of the polar parameter $\lambda_\vartheta$ in
one frame contains an ambiguity which prevents fundamental
(model-independent) interpretations of the results. The polarization
is only fully determined when \emph{both} the polar and the
azimuthal components of the decay distribution are known, or when
the distribution is analyzed in at least two geometrically
complementary frames.

\begin{figure}[tbh]
\centering
\includegraphics[width=0.495\linewidth]{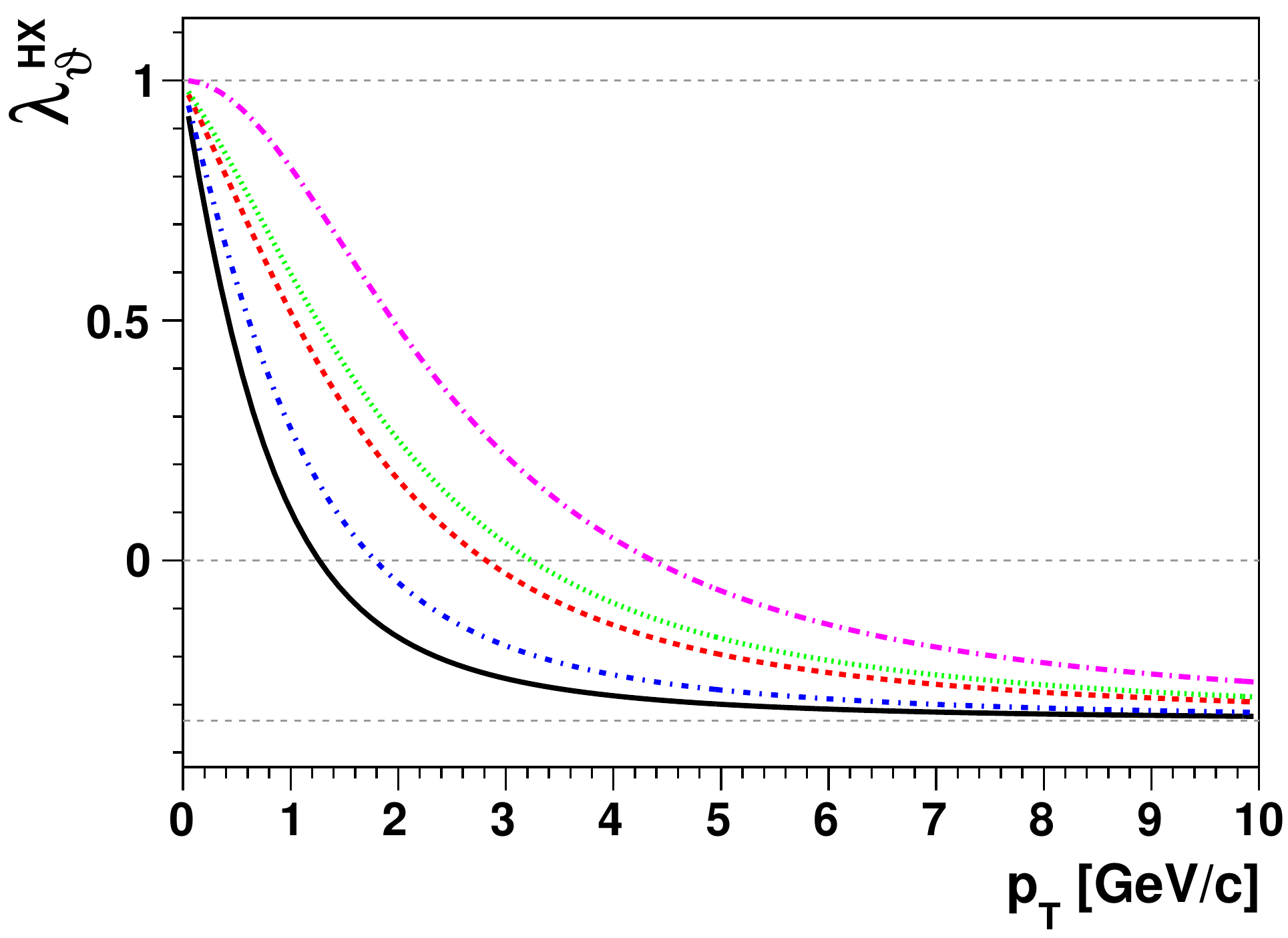}
\includegraphics[width=0.495\linewidth]{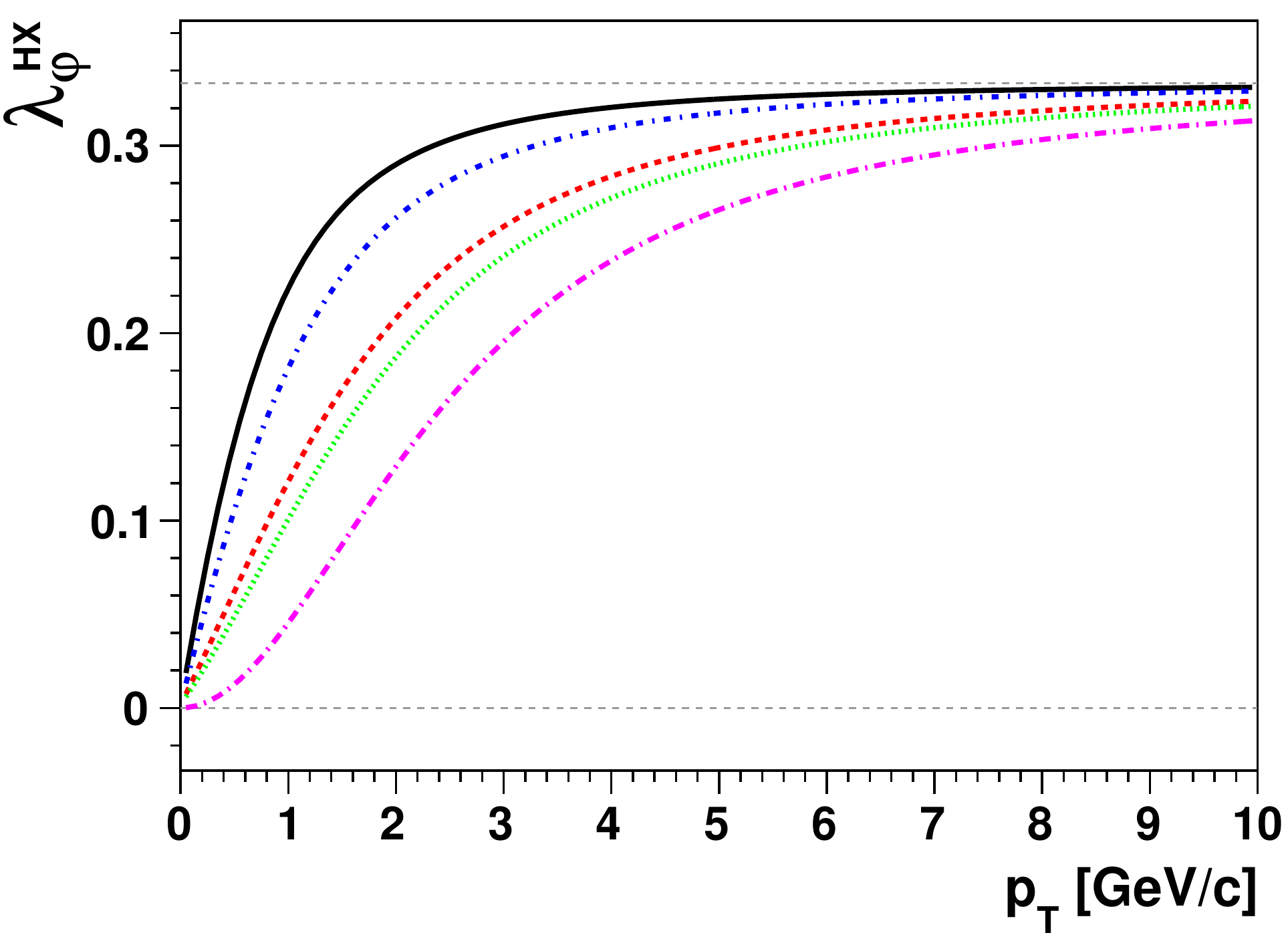}
\caption{Kinematic dependence of the \jpsi\ decay angular
  distribution seen in the HX frame, for a natural polarization $\lambda_{\vartheta}$\,$=$\,$+1$
  in the CS frame. The curves correspond to different rapidity intervals; from the solid line:
  $|y| < 0.6$ (CDF), $|y| < 0.9$ (ALICE), $|y| < 1.8$ (D0), $|y| <
  2.5$ (ATLAS and CMS), $2< y < 5$ (LHCb). For simplicity the event
  populations were generated flat in rapidity.}
\label{fig:kindep_lambda}
\end{figure}

Due to their frame-dependence, the parameters $\lambda_\vartheta$,
$\lambda_\varphi$ and $\lambda_{\vartheta \varphi}$ can be affected
by a strong \emph{explicit} kinematic dependence, reflecting the
change in direction of the chosen experimental axis (with respect to
the ``natural axis'') as a function of the quarkonium momentum. As
an example, we show in Fig.~\ref{fig:kindep_lambda} how a natural
transverse \jpsi\ polarization ($\lambda_\vartheta = +1$) in the CS
frame (with $\lambda_{\varphi} = \lambda_{\vartheta \varphi} = 0$
and no intrinsic kinematic dependence) translates into different
\pt-dependent polarizations measured in the HX frame in different
rapidity acceptance windows, representative of the acceptance ranges
of several Tevatron and LHC experiments.
This example shows that an ``unlucky'' choice of the observation
frame may lead to a rather misleading representation of the
experimental result. Moreover, the strong kinematic dependence
induced by such a choice may mimic and/or mask the fundamental
(``intrinsic'') dependencies reflecting the production mechanisms.

\begin{figure}[tbh]
\centering
\includegraphics[width=0.495\linewidth]{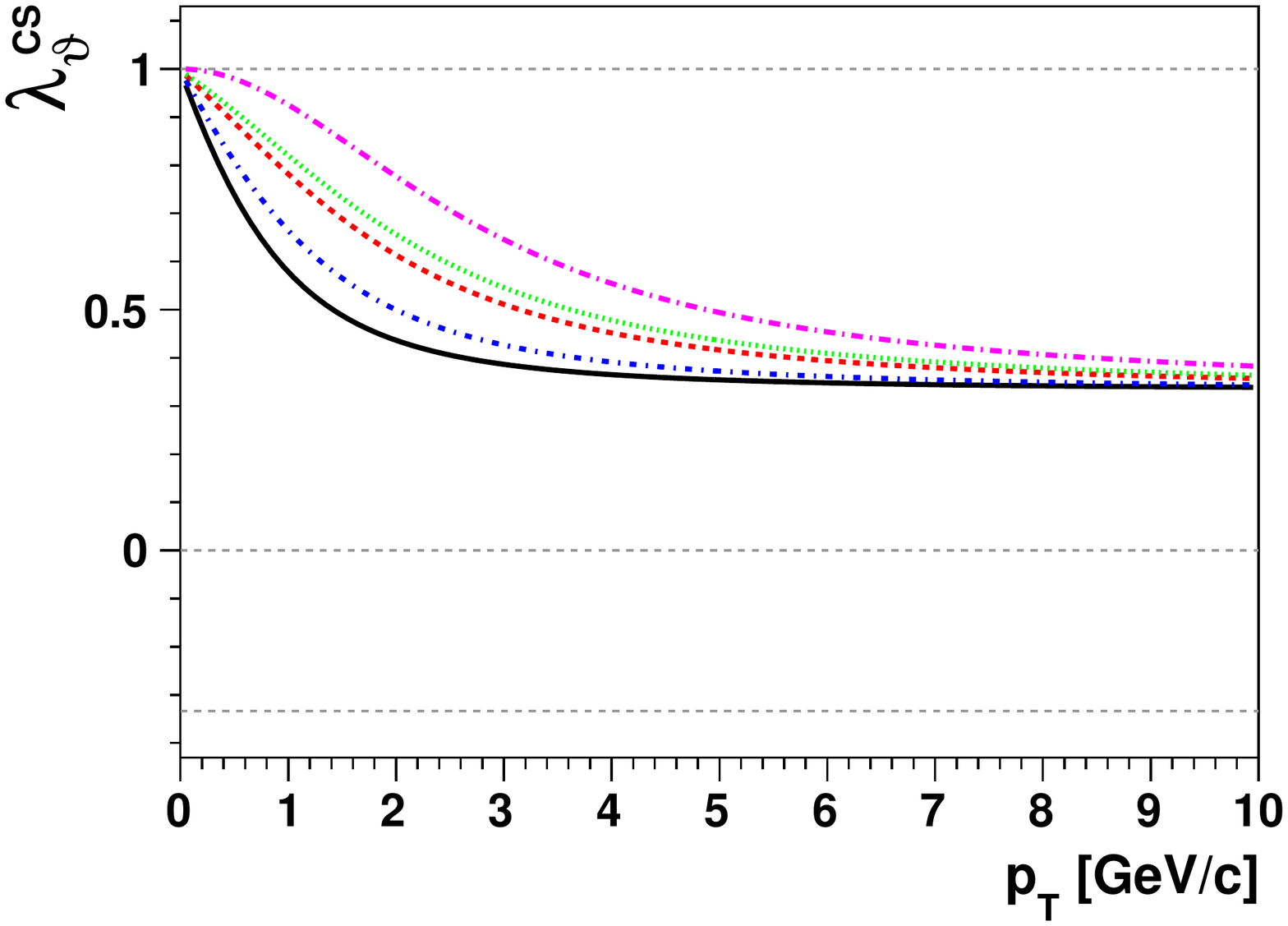}
\includegraphics[width=0.495\linewidth]{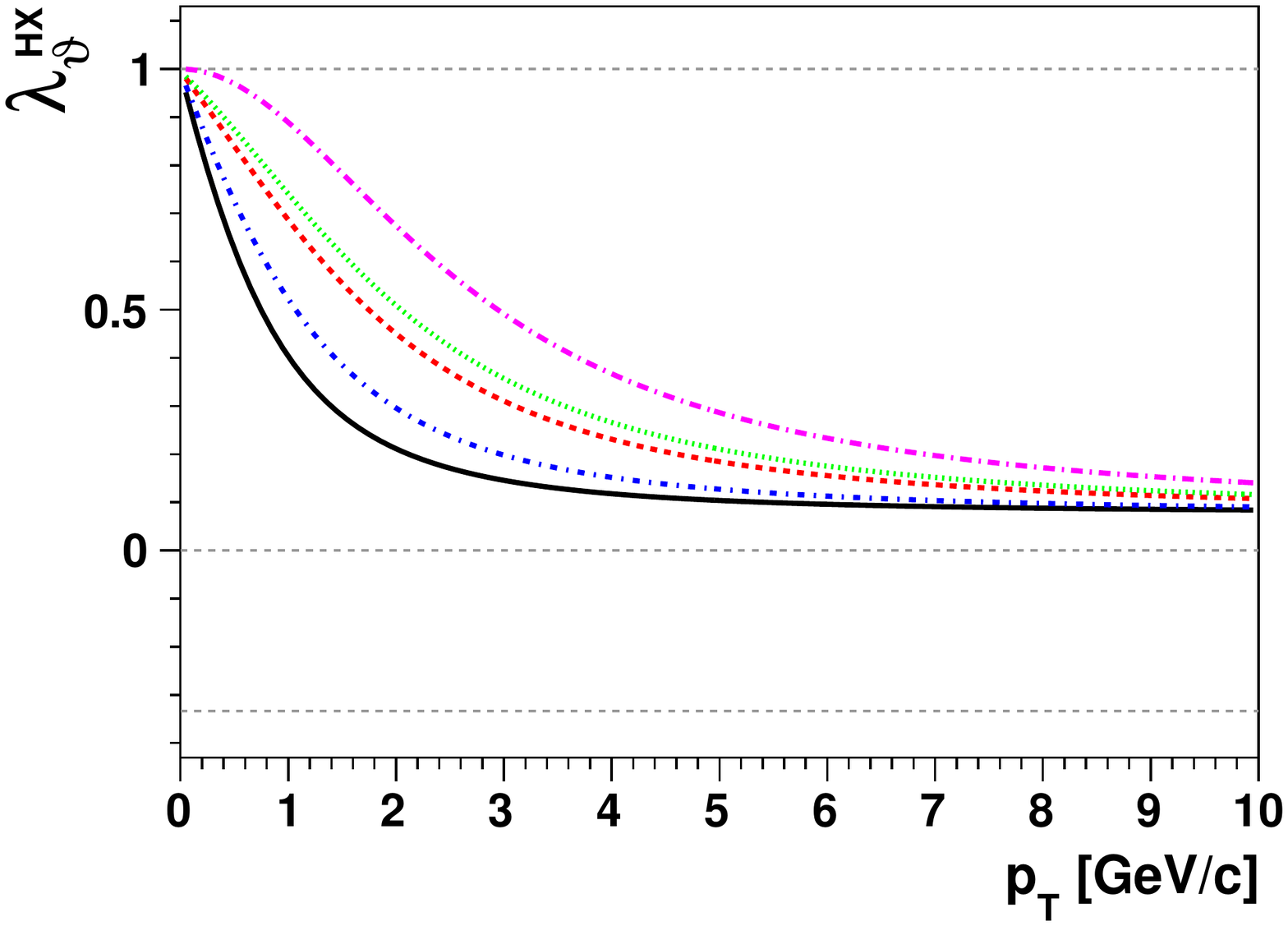}
\caption{Polar anisotropy of the \jpsi\ decay distribution as seen in the CS
(left) and HX (right) frames, when all the events have full transverse
polarization, but 60\% in the CS frame and 40\% in the HX frame. The curves
represent measurements in different rapidity
  ranges (see Fig.~\ref{fig:kindep_lambda}). } \label{fig:kindep_lambda_mix}
\end{figure}
Not always an ``optimal'' quantization axis exists. This is shown in
Fig.~\ref{fig:kindep_lambda_mix}, where we consider, for
illustration, that $60\%$ of the \jpsi\ events have natural
polarization $\lambda_\vartheta = +1$ in the CS frame while the
remaining fraction has $\lambda_\vartheta = +1$ in the HX frame.
Although the polarizations of the two event subsamples are
intrinsically independent of the production kinematics, in neither
frame, CS or HX, will measurements performed in different transverse
and longitudinal momenta windows find ``simple'', identical results.
Corresponding figures for the \upsOneS\ case can be seen in
Ref.~\refcite{bib:ImprovedQQbarPol}.

%
\begin{figure}[tbh]
\centering
\includegraphics[width=0.55\linewidth]{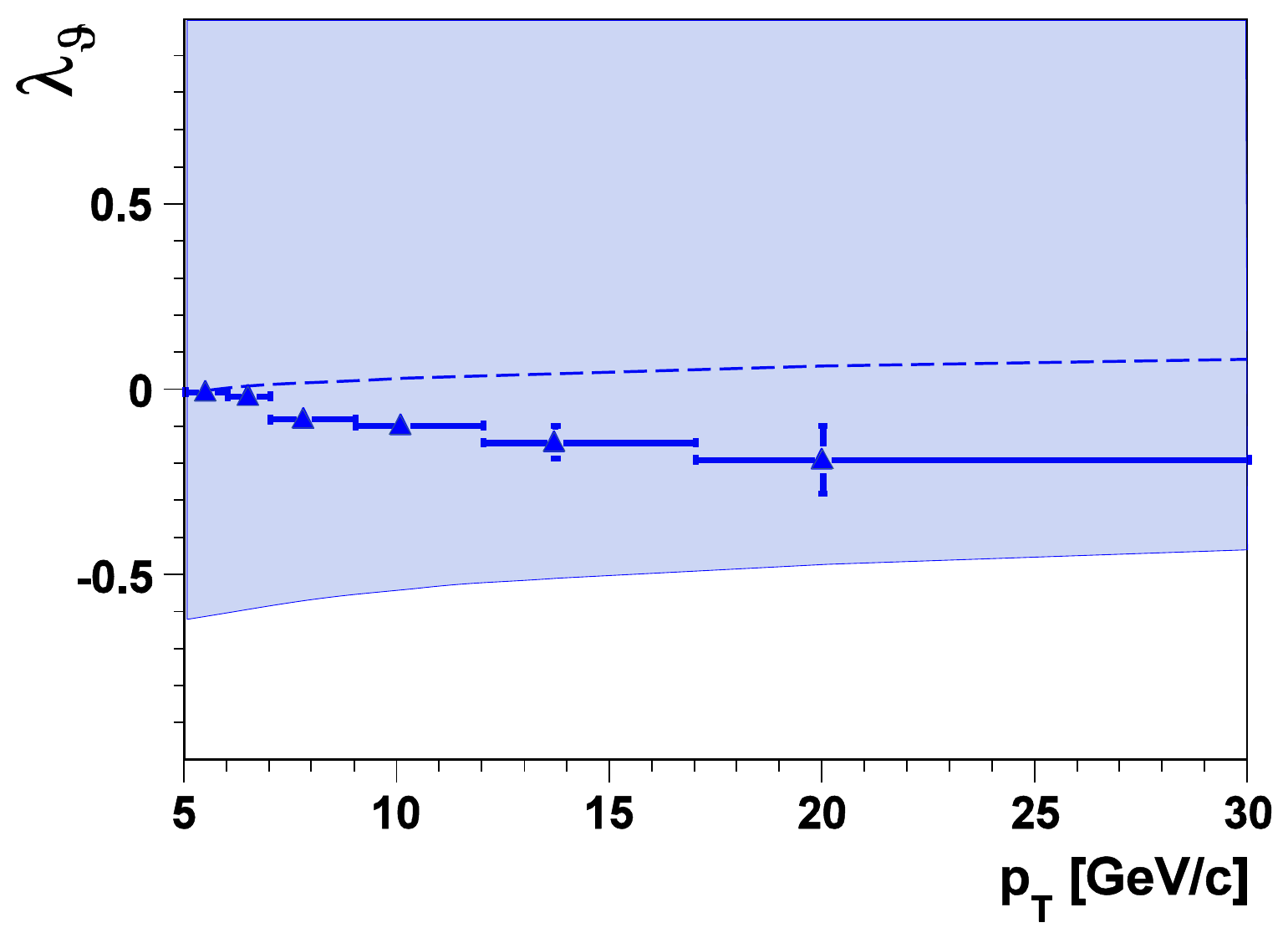}
\caption{The CDF \jpsi\ polarization measurement in the helicity frame (data
points) and the range for the corresponding polarization in the CS frame
(dashed line: CS polarization for $\lambda_\varphi^{\rm{HX}}=0$). }
\label{fig:CDF}
\end{figure}
%
CDF measured for the \jpsi\ an almost vanishing polar anisotropy parameter in
the helicity frame. It is natural to wonder how the measurement would look like
in the CS frame. However, the transformation to another frame depends on the
azimuthal anisotropy, which was not reported by the experiment. For example, as
shown in Fig.~\ref{fig:CDF}, if the distribution in the HX frame were
azimuthally isotropic, the measured polarization would correspond to a
practically undetectable polarization in the CS frame (dashed line). However,
if we take into account all physically possible values of the azimuthal
anisotropy, as allowed by the relations in Eq.~\ref{eq:triangles}, we can only
derive a broad spectrum of possible CS polarizations, approximately included
between $-0.5$ and $+1$ (shaded band). This example shows how a measurement
reporting only the polar anisotropy is amenable to several interpretations in
fundamental terms, often corresponding to drastically different physical cases.

\begin{figure}[tbh]
\centering
\includegraphics[width=0.98\linewidth]{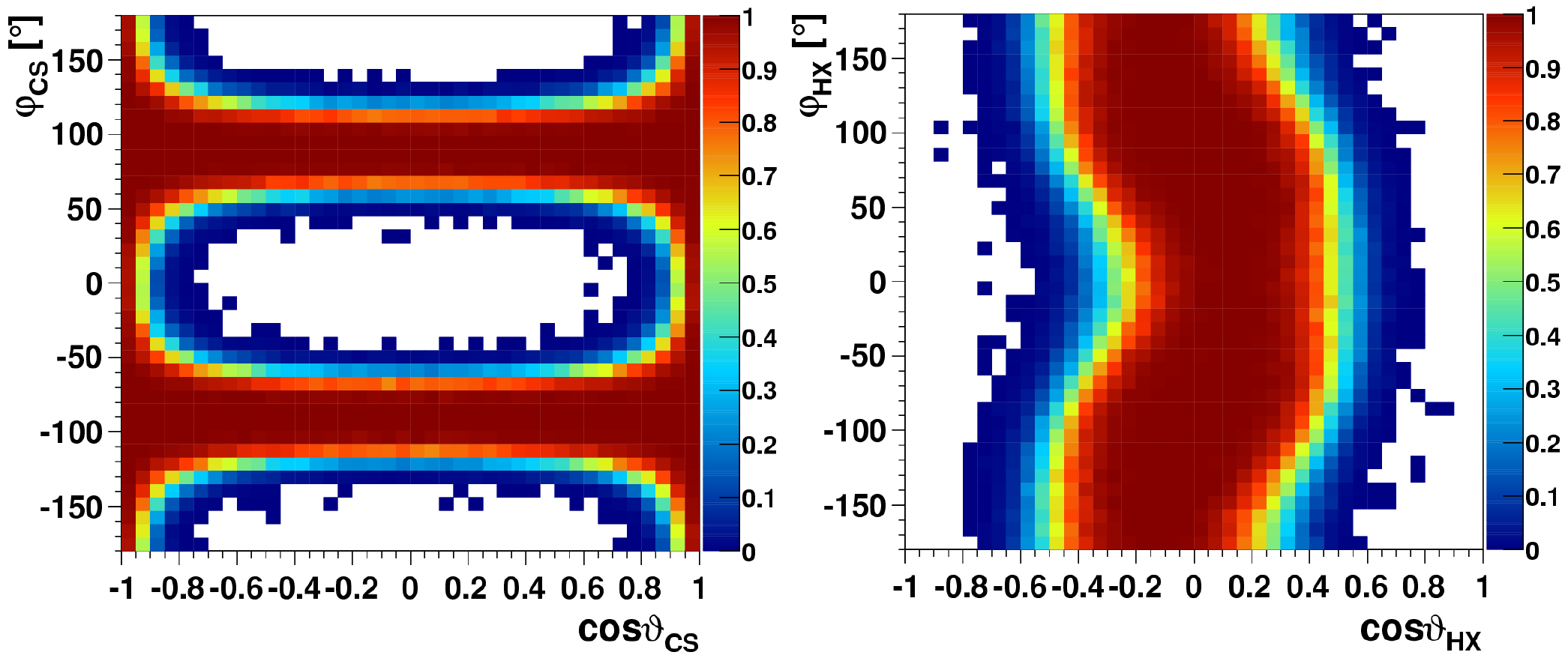}
\caption{Angular acceptances in the CS and HX frames for \jpsi\ decay into
leptons, in the kinematic region $9 < p_{\rm T} < 12$~GeV$/c$, $0 < |y| < 1$,
when only leptons having $p_{\rm T} > 3$~GeV$/c$ are detected.}
\label{fig:acceptances}
\end{figure}

An analysis ignoring the azimuthal dimension can produce \emph{wrong} results.
In fact, the experimental acceptances for the variables $\cos\vartheta$ and
$\varphi$ are usually strongly intercorrelated because of the limited
sensitivity to low-momentum leptons, which reduces the population of events in
specific angular regions, depending on the reference frame. For example, the
experimental efficiency for the projected $\cos\vartheta$ distribution depends
on the $\varphi$ distribution, that is on $\lambda_\varphi$, and vice-versa. If
the $\varphi$ dimension is integrated out and ignored, the $\lambda_\vartheta$
measurement becomes strongly dependent on the specific ``prior hypothesis''
(implicitly) made for the angular distribution in the Monte Carlo simulation.
To illustrate this concept, we consider \jpsi\ pseudo-data in the kinematic
region $9 < p_{\rm T} < 12$~GeV$/c$, $0 < |y| < 1$, simulating the acceptance
filter with the requirement that both leptons have $p_{\rm T} > 3$~GeV$/c$. The
angular acceptances for these conditions in the CS and HX frames are shown in
Fig.~\ref{fig:acceptances}.
%
%
%
\begin{figure}[p]
\centering
\includegraphics[width=0.778\linewidth]{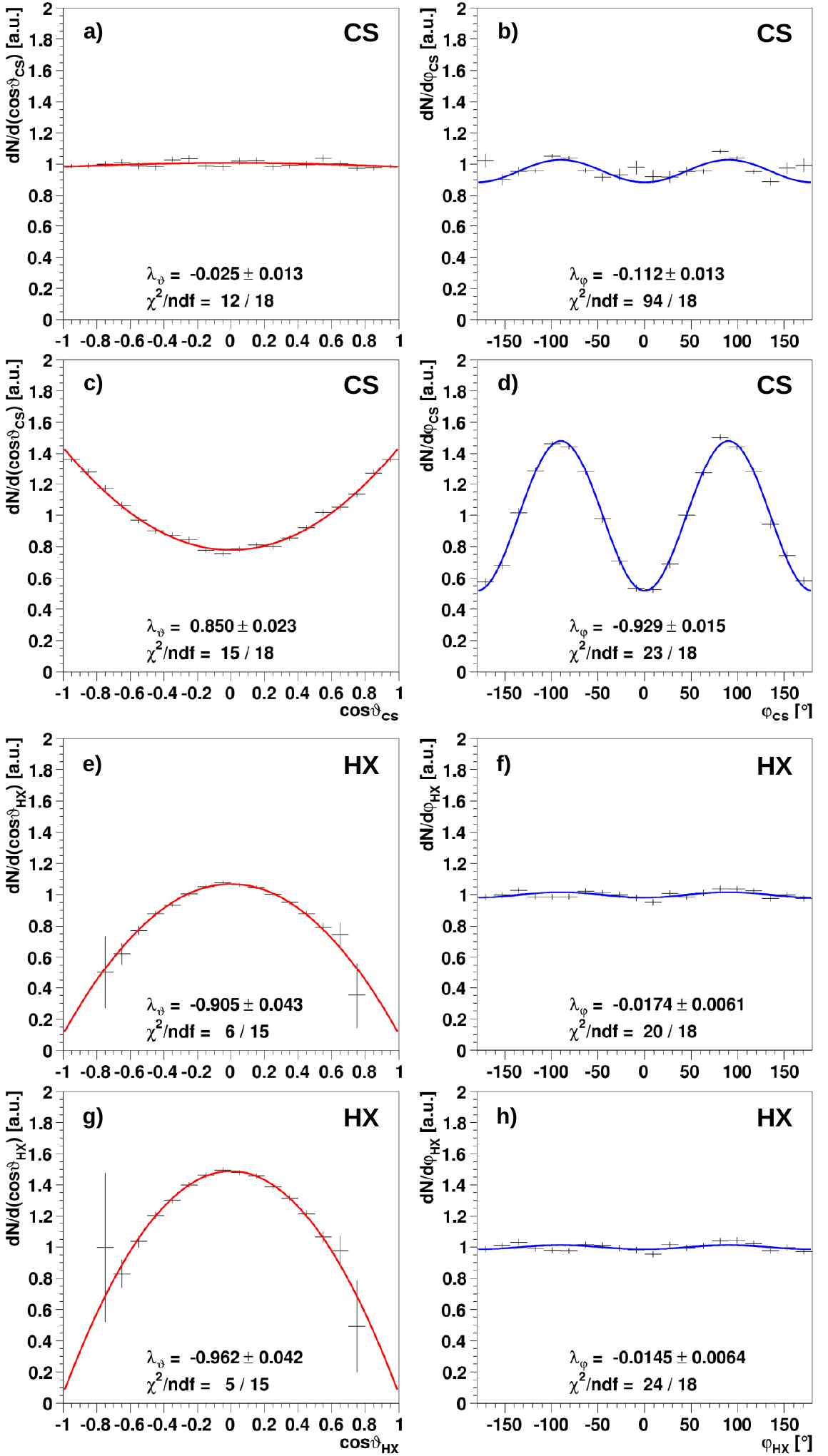}
\caption{Results of a pseudo-experiment ($\sim 60$k reconstructed dilepton
  events) where the \jpsi\ polarization (generated as fully longitudinal in the HX frame) is
  measured through one-dimensional angular projections in the CS and HX frames. a,b,e,f: (wrong) results obtained
  using a ``standard'' unpolarized Monte Carlo simulation for the acceptance correction.
  c,d,g,h: (``correct'') results obtained after reweighing iteratively the Monte Carlo
  data according to the results found.} \label{fig:mistakeAccCorr}
\end{figure}
%
We consider the example scenario of a fully longitudinal polarization in the HX
frame. A one-dimensional measurement is performed in the CS frame integrating
out and ignoring the $\varphi$ dependence. The detector-acceptance correction
is performed one-dimensionally, using Monte Carlo data generated assuming a
flat azimuthal dependence. Figure~\ref{fig:mistakeAccCorr}a shows that the
acceptance-corrected $\cos\vartheta$ distribution in the CS frame is flat,
leading to a wrong ``unpolarized'' result, reflecting the polarization
assumption used in the Monte Carlo simulation. If the Monte Carlo data used for
the acceptance correction are reweighted to the ``true'' polarization (a
\emph{two-dimensional} ingredient), the same distribution changes drastically
(Fig.~\ref{fig:mistakeAccCorr}c), correctly showing a strong transverse
polarization (the CS frame being almost perpendicular to the HX frame). This
shows that when only a one-dimensional projected distribution is measured, the
detector acceptance description must, nevertheless, be maintained
multi-dimensional. One-dimensional acceptance corrections or ``template'' fits
should be avoided, unless the MC is iteratively re-generated with the correct
distribution of the variables that have been integrated out (which has,
therefore, to be measured anyway). Unfortunately, one-dimensional polarization
analyses are widespread, even, paradoxically, in precision tests of the
Standard Model and searches for new physics. To measure the polar anisotropy of
$W$ decays or Drell--Yan production using template distributions (to account
for acceptance and efficiency) integrated over the azimuthal angle, as done in
recent analyses reported by LHC experiments, may strongly bias the measurement
towards the distribution used to produce the Monte Carlo simulation. One
analysis even \emph{imposes} the absence of azimuthal anisotropies, assuming
that the data are exactly described by the naive Born-level Drell--Yan angular
distribution valid at $p_{\rm T} = 0$. New physics effects changing drastically
the azimuthal anisotropy with respect to the expected one (assumed in the Monte
Carlo) may be missed by this kind of analyses.

\section{A frame-invariant approach}
\label{sec:invariant}

It can be shown that the combination of coefficients
\begin{equation}
  \tilde{\lambda} \, = \, (\lambda_\vartheta + 3 \lambda_\varphi)/(1 -
    \lambda_\varphi) \label{eq:lambda_tilde}
\end{equation}
is independent of the polarization frame, The fundamental meaning of
the frame-invariance of this quantity is discussed in
Ref.~\refcite{bib:LTGen}. The determination of $\tilde{\lambda}$ is
immune to ``extrinsic'' kinematic dependencies induced by the
observation perspective and is, therefore, less acceptance-dependent
than the standard anisotropy parameters $\lambda_{\vartheta}$,
$\lambda_{\varphi}$ and $\lambda_{\vartheta \varphi}$. Referring to
the example shown in Fig.~\ref{fig:kindep_lambda_mix}, any arbitrary
choice of the experimental observation frame will always yield the
value $\tilde{\lambda} = +1$, independently of kinematics.
This particular case, where all contributing processes are
transversely polarized, is formally equivalent to the Lam-Tung
relation.\cite{bib:LamTung} The existence of frame-invariant
parameters also provides a useful instrument for experimental
analyses. Checking, for example, that the same value of an invariant
quantity is obtained, within systematic uncertainties, in two
distinct polarization frames is a non-trivial verification of the
absence of unaccounted systematic effects.
In fact, detector geometry and/or data selection constraints strongly polarize
the reconstructed dilepton events, as shown in Fig.~\ref{fig:acceptances}.
Background processes also affect the measured polarization, if not well
subtracted.
The spurious anisotropies induced by detector effects and background
do not obey the frame transformation rules characteristic of a
physical $J=1$ state.
If not well corrected and subtracted, these effects will distort the
shape of the measured decay distribution differently in different
polarization frames. In particular, they will violate the
frame-independent relations between the angular parameters.
Any two physical polarization axes (defined in the rest frame of the meson and
belonging to the production plane) may be chosen to perform these ``sanity
tests''. The HX and CS frames are ideal choices at high \pt\ and mid rapidity,
where they tend to be orthogonal to each other.
At forward rapidity and low \pt, the significance of the test can be
maximized by using the CS axis and the ``perpendicular helicity
axis''~\cite{bib:perpHelicity}, which coincides with the helicity
axis at zero rapidity and remains orthogonal to the CS axis at
nonzero rapidity.
Given that $\tilde{\lambda}$ is ``homogeneous'' to the anisotropy
parameters, the difference $\tilde{\lambda}^{({\rm B})} -
\tilde{\lambda}^{({\rm A})}$ between the results obtained in two
frames provides a direct evaluation of the level of systematic
uncertainties not accounted in the analysis.

To illustrate the application of the frame-independent formalism as
a tool to spot problems in experimental data analyses, we refer
again to the above-described \jpsi\ pseudo-experiments. The result
of the measurement performed with one-dimensional acceptance
correction assuming unpolarized production is shown in
Fig.~\ref{fig:mistakeAccCorr}a,b,e,f, including, this time, both the
polar and azimuthal projections, in the CS and in the HX frame. From
the comparison of these four one-dimensional results we derive,
using Eq.~\ref{eq:lambda_tilde}, $\tilde{\lambda}^{\rm CS} \simeq
-0.32$ and $\tilde{\lambda}^{\rm HX} \simeq -0.93$. The difference
between the two values is an unequivocal signal of a mistake in the
analysis. In fact, after reweighting the Monte Carlo with the
``correct'' polarization (Fig.~\ref{fig:mistakeAccCorr}c,d,g,h),
which can be iteratively inferred choosing as ``generation frame''
the one showing at each step the strongest polarization modulations
(the HX frame in this case), both $\tilde{\lambda}$ values approach
$-1$, as expected in this exercise.

Incidentally, we note that the ``true'' $\tilde{\lambda}$ value is
generally not included between the values found in two different
reference frames. The best value of $\tilde{\lambda}$ in the
presence of not completely corrected systematic effects is not the
average among different frames and is best approximated by the value
found in the frame showing the smallest acceptance correlations
between $\cos\vartheta$ and $\varphi$ (HX frame in the above
example, see Fig.~\ref{fig:acceptances}). It is, therefore, a priori
not fully justified to impose the constraint $\tilde{\lambda}^{\rm
CS} = \tilde{\lambda}^{\rm HX}$ in a fit of the angular
distributions performed simultaneously in two frames, as done in a
recent LHC analysis of \jpsi\ polarization.

%
\begin{figure}[tbh]
\centering
\includegraphics[width=0.42\linewidth]{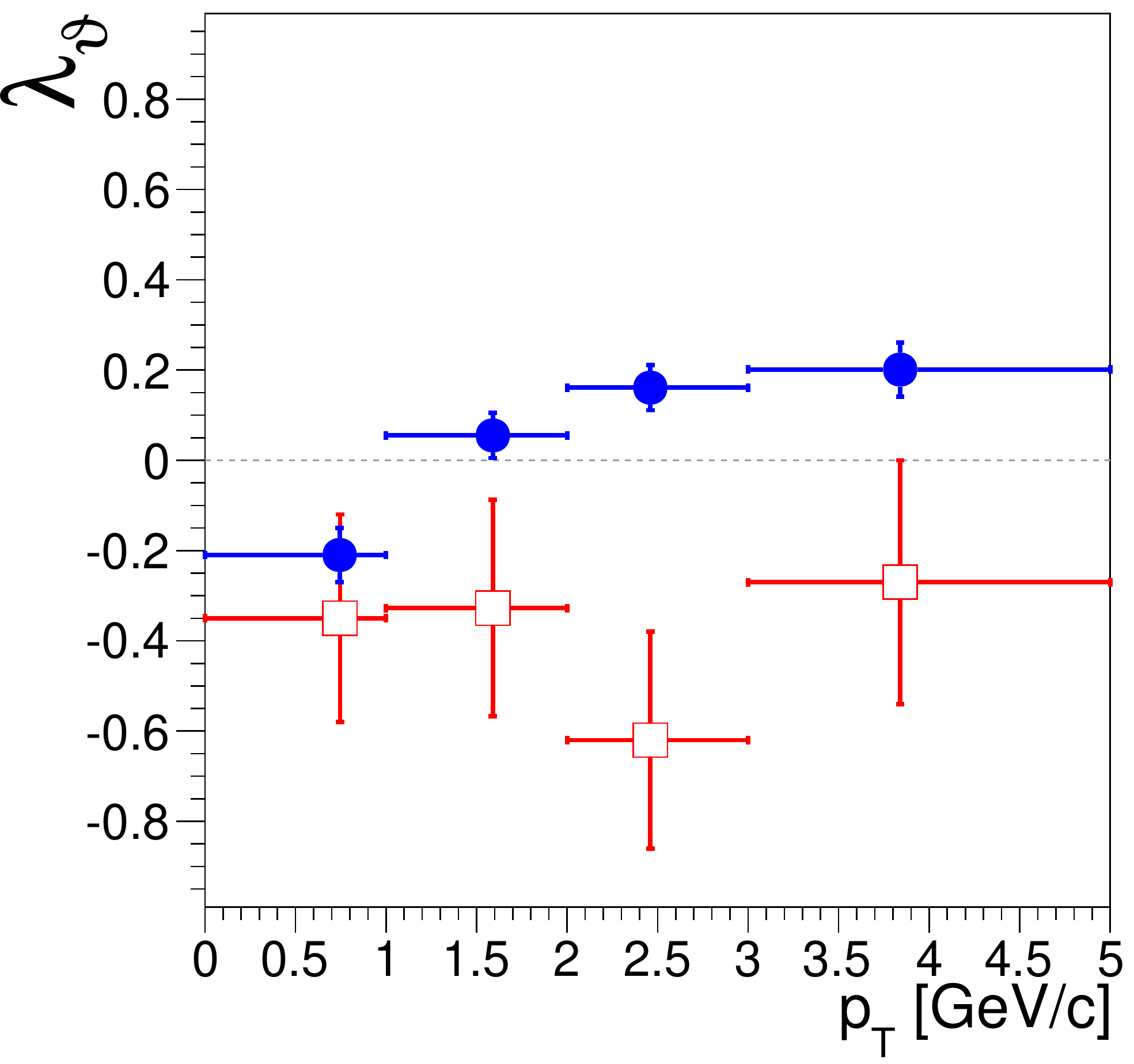}
\includegraphics[width=0.42\linewidth]{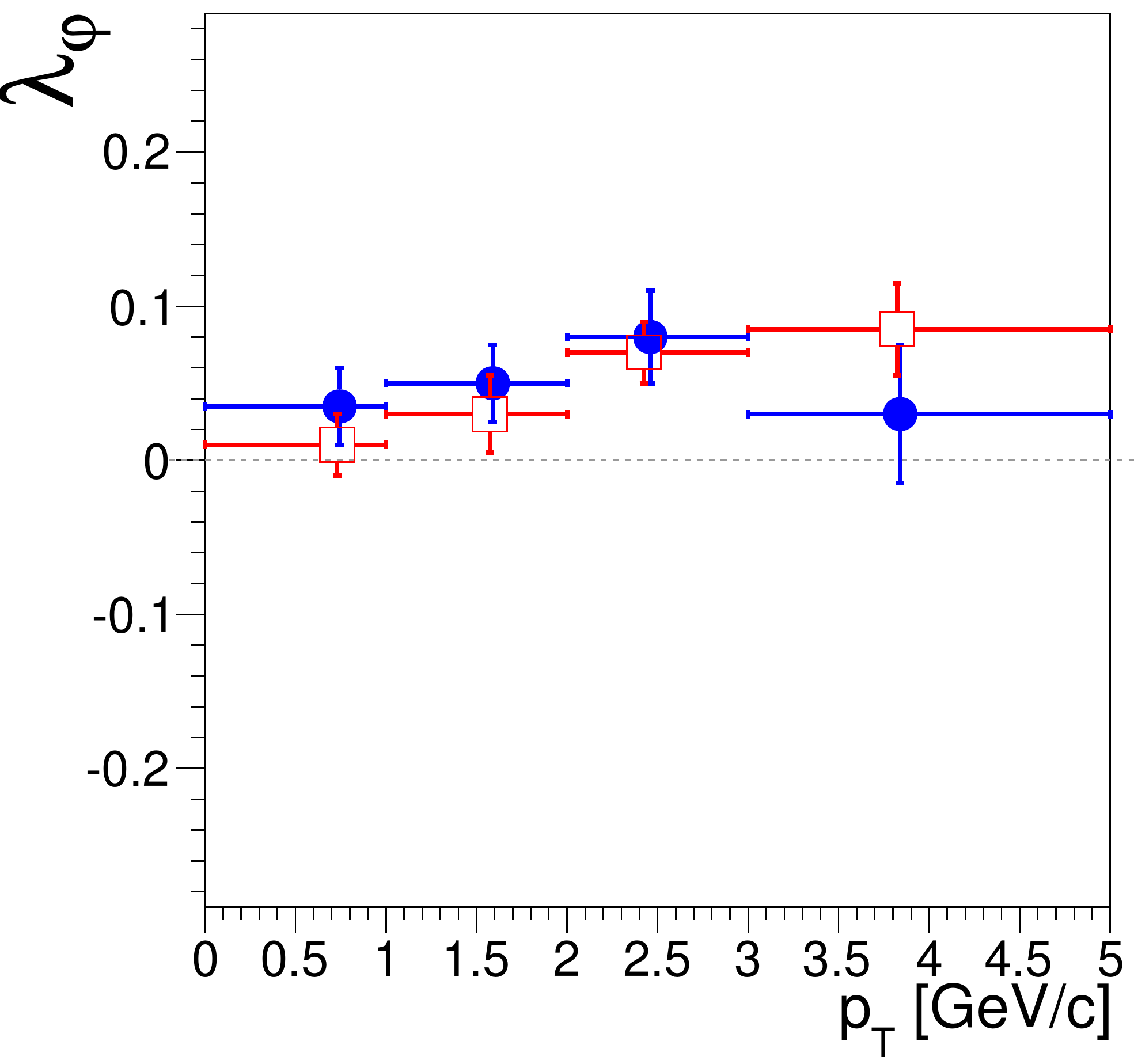}
\caption{Example of data where the \jpsi\
  polarization measurements in the CS and HX frames (empty and filled
  symbols, respectively) are inconsistent with each other.}
\label{fig:NA60}
\end{figure}
%
Another example of utility of the invariant polarization parameter can be seen
in Fig.~\ref{fig:NA60}, showing \jpsi\ polarization ``measurements'' in the CS
and HX frames versus \pt. While the $\lambda_\vartheta$ values seem to change
significantly from one frame to the other, the two $\lambda_\varphi$ patterns
are very similar.
This observation alerts to an experimental artifact in the data
analysis.  We can evaluate the significance of the contradiction by
calculating the frame-invariant $\tilde{\lambda}$ variable in each
of the two frames.  For the case illustrated in Fig.~\ref{fig:NA60},
averaging the four represented \pt\ bins, we see that
$\tilde{\lambda}$ in the HX frame is larger than in the CS frame by
0.5 (a rather large value, considering that the decay parameters are
bound between $-1$ and $+1$).  In other words, the determination of
the decay parameters must be biased by systematic errors of roughly
this magnitude.
Given the puzzles and contradictions existing in the published
experimental results, as recalled in Section~\ref{sec:exp}, the use
of a frame-invariant approach to perform self-consistency checks,
which can expose unaccounted systematic effects due to detector
limitations and analysis biases, constitutes a non-trivial
complementary aspect of the methodologies for quarkonium
polarization measurements.

\section{The role of the feed-down decays}
\label{sec:feeddown}

Many of the prompt J/$\psi$ and $\Upsilon$ mesons produced in hadronic
collisions result from the decay of heavier $S$- or $P$-wave quarkonia.
However, the existing polarization measurements at collider energies make no
distinction between directly and indirectly produced states. The role of the
feed-down from heavier $S$ states (responsible, for example, for about $8\%$ of
J/$\psi$ production at low $p_{\rm T}$\cite{bib:feeddown}) is rather well
understood. Data of the BES\cite{bib:BES_psiprime} and
CLEO\cite{bib:CLEO_upsilon1,bib:CLEO_upsilon2} experiments in $e^+e^-$
collisions indicate that in the decays $\psi^{\prime} \rightarrow
\mathrm{J}/\psi \pi \pi$ and $\Upsilon(2S) \rightarrow \Upsilon(1S) \pi \pi$
the di-pion system is produced predominantly in the spatially isotropic
($S$-wave) configuration, meaning that no angular momentum is transferred to
it. Consequently, the angular momentum alignment is preserved in the transition
from the $2S$ to the $1S$ state. This allows us to assume that the dilepton
decay angular distribution of the J/$\psi$ [$\Upsilon(1S)$] mesons resulting
from $\psi^{\prime}$ [$\Upsilon(2/3S)$] decays is the same as the one of the
$\psi^{\prime}$ [$\Upsilon(2/3S)$], provided that a common polarization axis is
chosen for the two particles. At high momentum, when the J/$\psi$ and
$\psi^{\prime}$ directions with respect to the centre of mass of the colliding
hadrons practically coincide, $\psi^{\prime}$ mesons and J/$\psi$ mesons from
$\psi^{\prime}$ decays have the same observable polarization with respect to
any system of axes defined on the basis of the directions of the colliding
hadrons. In the case of the polar anisotropy parameter $\lambda_\vartheta$, for
instance, the relative error, $|\Delta\lambda_\vartheta/\lambda_\vartheta|$,
induced by the approximation of considering the J/$\psi$ and $\psi^{\prime}$
directions as coinciding is $\mathcal{O}[(\Delta m / p)^2]$, where $\Delta m$
is the $2S-1S$ mass difference and $p$ the total laboratory momentum of the
dilepton. For $p>5$~GeV/$c$ this error is of order 1\%. Moreover, the directly
produced J/$\psi$ [$\Upsilon(1S)$] and $\psi^\prime$ [$\Upsilon(2/3S)$] are
expected to have the same production mechanisms and, therefore, very similar
polarizations. As a consequence, the polarization of J/$\psi$ [$\Upsilon(1S)$]
from $\psi^\prime$ [$\Upsilon(2/3S)$] can be considered to be almost equal to
the polarization of directly produced J/$\psi$ [$\Upsilon(1S)$], so that, at
least in first approximation, the two contributions can be treated as one.

On the contrary, the J/$\psi$ [$\Upsilon(1S)$] mesons resulting from
$\chi_{cJ}$ [$\chi_{bJ}$] radiative decays can have very different
polarizations with respect to the directly produced ones. Directly produced $P$
and $S$ states can originate from different partonic and long-distance
processes, given their different angular momentum and parity properties.
Moreover, the emission of the spin-1 and always transversely polarized photon
necessarily changes the angular momentum projection of the $q \bar{q}$ system
in the $P \to S$ radiative transition. As a result, the relation between the
``spin-alignment'' of the directly produced $P$ or $S$ state and the shape of
the observed dilepton angular distribution is totally different in the two
cases: for example, if directly produced J/$\psi$, $\chi_{c1}$ and $\chi_{c2}$
all had ``longitudinal'' polarization (angular momentum projection $J_z = 0$
along a given quantization axis), the shape of the dilepton distribution would
be of the kind $1 - \cos^2\!\vartheta$ for the direct J/$\psi$, $1 +
\cos^2\!\vartheta$ for the J/$\psi$ from $\chi_{c1}$ and $1 -\frac{3}{5}
\cos^2\!\vartheta$ for the J/$\psi$ from $\chi_{c2}$. While for directly
produced $S$ states $-1 < \lambda_\vartheta < +1$, for those from decays of
$P_1$ and $P_2$ states the lower bound is $-1/3$ and $-3/5$, respectively. More
detailed constraints on the three anisotropy parameters $\lambda_\vartheta$,
$\lambda_\varphi$ and $\lambda_{\vartheta \varphi}$ in the cases of directly
produced $S$ state and $S$ states from decays of $P_1$ and $P_2$ states can be
found in Ref.~\refcite{bib:chi_polarization}. Figure 3 of that work shows that
the allowed parameter space of the decay anisotropy parameters for the directly
produced J/$\psi$ [$\Upsilon(1S)$] strictly includes the one of the $S$-states
from $P_2$ decays, which, in turn, strictly includes the one of the $S$-states
from $P_1$ decays.

The feed-down fractions are not well-known experimentally. In the charmonium
case, the $\chi_{c}$-to-J/$\psi$ and $\chi_{c2}$-to-$\chi_{c1}$ yield ratios
have been measured by CDF\cite{bib:cdf_chic1,bib:cdf_chic2} in the rapidity
interval $|y|< 0.6$, with insufficient precision to indicate or exclude
important $p_{\rm T}$ dependencies. The $p_{\rm T}$-averaged results,
\begin{align}
\begin{split}
R(\chi_{c1}) + R(\chi_{c2}) & = 0.30 \pm 0.06 \, , \\
\quad R(\chi_{c2}) / R(\chi_{c1}) & = 0.40 \pm 0.02 \, ,
\label{eq:Rchic}
\end{split}
\end{align}
where $R(\chi_{c1})$ and $R(\chi_{c2})$ are the fractions of prompt
J/$\psi$ yield due to the radiative decays of $\chi_{c1}$ and
$\chi_{c2}$, effectively correspond to a phase-space region (low
$p_{\rm T}$ and central rapidity), much smaller than the one covered
by the LHC experiments.

CDF also measured\cite{bib:cdf_chib} the fractions of $\Upsilon(1S)$
mesons coming from radiative decays of $1P$ and $2P$ states as,
respectively, $R(\chi_{b1})+R(\chi_{b2}) = (27 \pm 8)\%$ and
$R(\chi_{b1}^\prime) + R(\chi_{b2}^\prime) = (11 \pm 5)\%$, for
$p_{\rm T} > 8$~GeV$/c$ and without discrimination between the $J=1$
and $J=2$ states. These results tend to indicate that the
contribution of the feed-down from $P$ states to $\Upsilon(1S)$
production is at least as large as in the corresponding charmonium
case, even if the experimental error is quite large. The same
indication is provided with higher significance by the $\Upsilon$
polarization measurement of E866\cite{bib:e866_Ups}, at low $p_{\rm
T}$, as discussed below.

Using available experimental and theoretical information, we can derive two
illustrative scenarios for the polarizations of the charmonium and bottomonium
families.
\begin{figure}[tbh]
  \centering
  \includegraphics[width=0.59\textwidth]{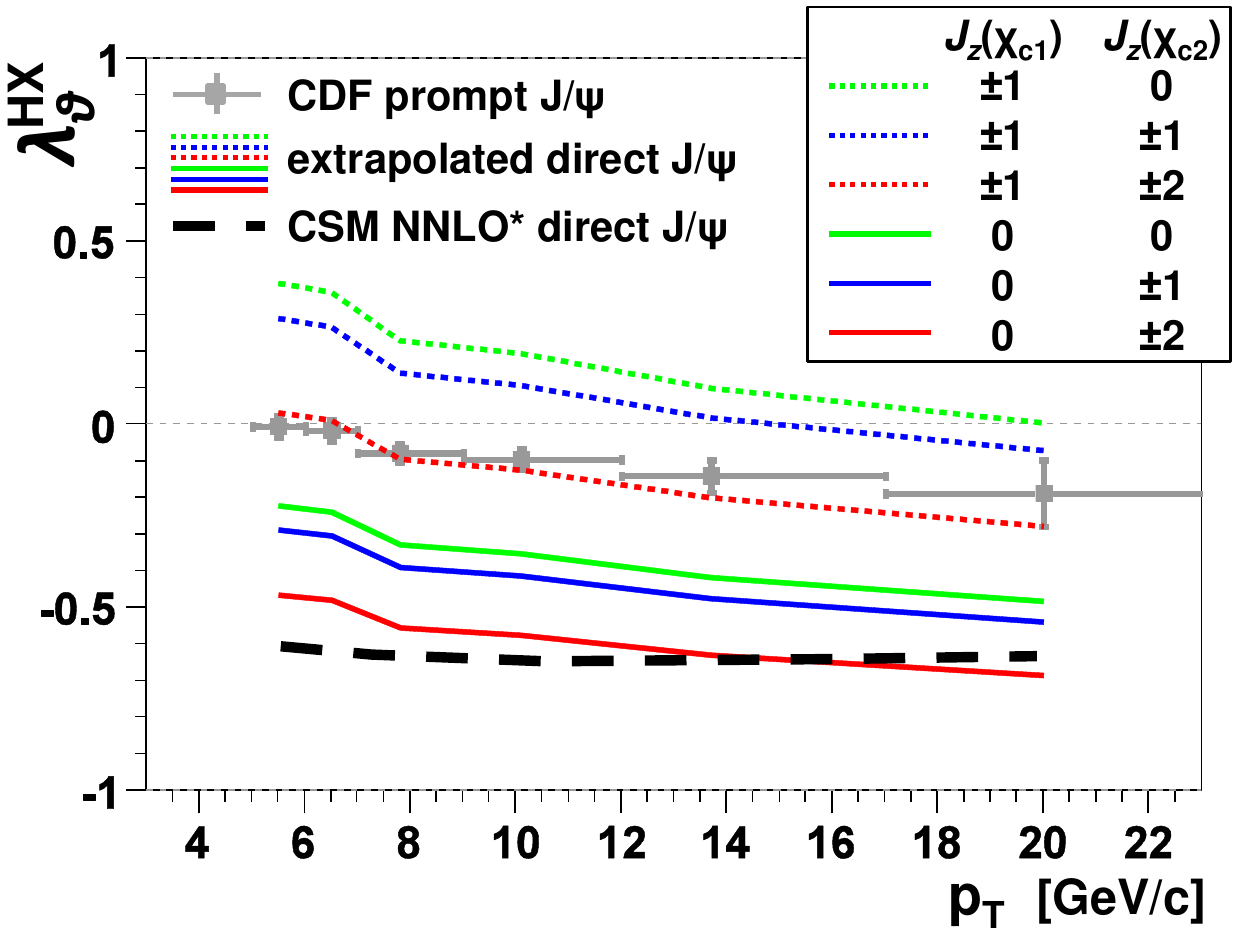}
  \caption{ Direct-J/$\psi$ polarizations ($\lambda_{\vartheta}$) extrapolated
  from the CDF measurement of prompt-J/$\psi$ polarization (in the helicity
  frame), using several scenarios for the $\chi_c$ polarizations. }
  \label{fig:CDF_direct_extrapolation}
\end{figure}
Figure~\ref{fig:CDF_direct_extrapolation} illustrates how the CDF measurement
of prompt-J/$\psi$ polarization\cite{bib:CDFpol2} can be translated in a range
of possible values of the direct-J/$\psi$ polarization, using the available
information about the feed-down fractions and all possible combinations of
hypotheses of pure polarization states for $\chi_{c1}$ and $\chi_{c2}$. The
feed-down fraction is set to 0.42, two standard deviations higher than the
central CDF value (Eq.~\ref{eq:Rchic}); using 0.30 simply decreases the spread
between the curves. The $R(\chi_{c2}) / R(\chi_{c1})$ ratio is set to 0.40;
changes remaining compatible with the CDF measurement give almost identical
curves. In the scenario in which $\chi_{c1}$ and $\chi_{c2}$ are produced with,
respectively, $J_z = 0$ and $J_z = \pm 2$ polarizations the CDF measurement is
seen to be described by partial next-to-next-to-leading order (NNLO$^*$) CSM
predictions for directly produced
$S$-states\cite{bib:CSM_directQQpol_Jpsi,bib:CSM_directQQpol_Y}. The validity
of this J/$\psi$ polarization scenario can be probed by experiments able to
discriminate if the J/$\psi$ is produced together with a photon such that the
two are compatible with being $\chi_{c1}$ or $\chi_{c2}$ decay products. Such
dilepton events, resulting from $\chi_c$ decays, should show a full transverse
polarization ($\lambda_\vartheta^{\chi_{c1}} = \lambda_\vartheta^{\chi_{c2}} =
+1$), while the directly produced J/$\psi$ mesons should have a strong
longitudinal polarization ($\lambda_\vartheta^\mathrm{dir} \simeq -0.6$).

\begin{figure}[tbh]
  \centering
  \includegraphics[width=0.59\textwidth]{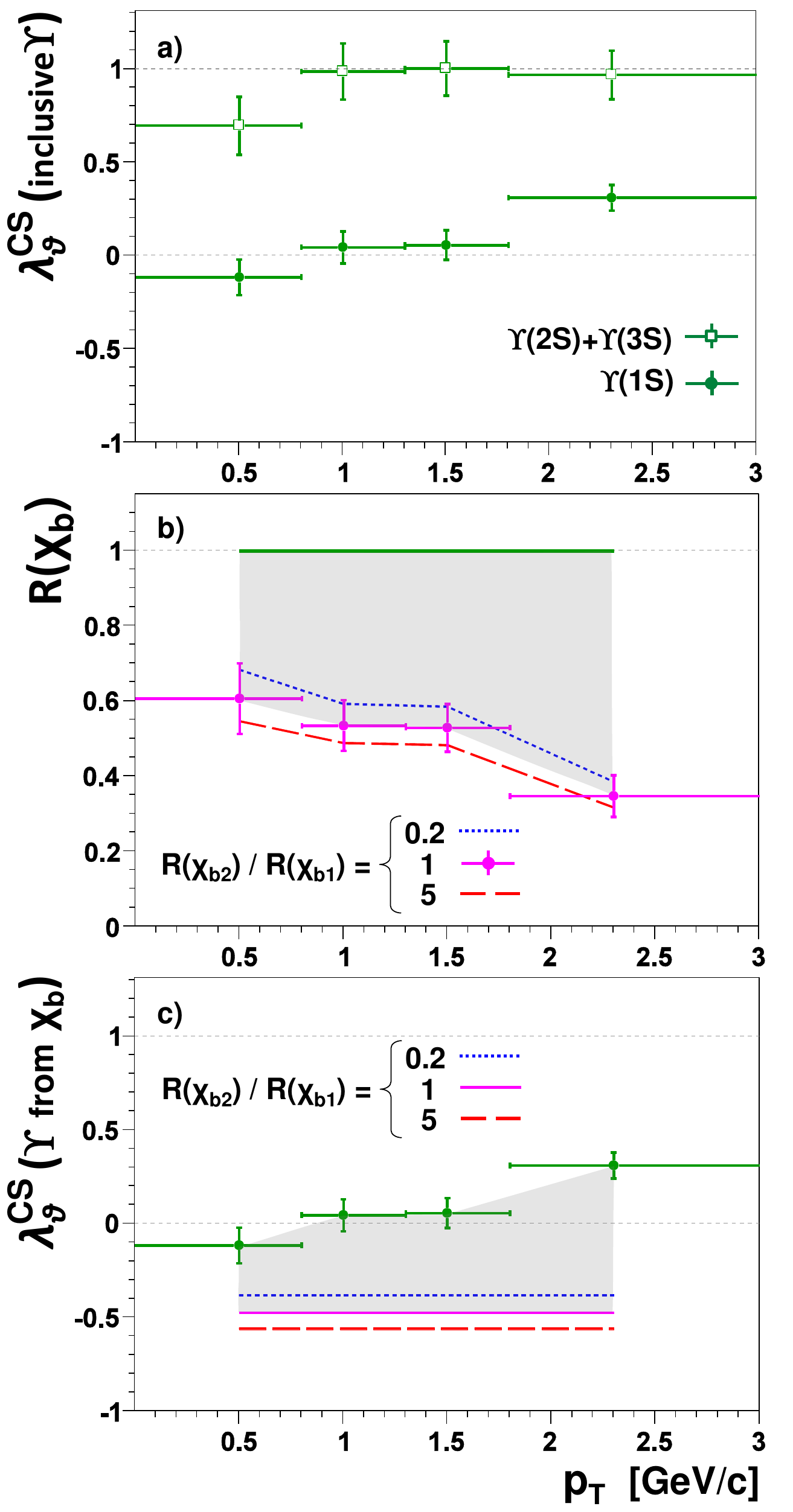}
  \caption{ The E866 measurement of $\Upsilon$
  polarizations in the CS frame as a function of $p_{\rm T}$ (a), the deduced ranges for the fraction of $\Upsilon(1S)$
  mesons coming from $\chi_{b}$ decays (b) and the deduced range of their possible polarizations (c).
  A systematic uncertainty of $\pm 0.06$ is not included in the error bars
  of the data points in (a).
  The error bars in the derived lower limit for $R(\chi_{b})$ (b) reflect the uncertainty
  in the $\lambda_{\vartheta}$ measurements, assuming that the global systematic uncertainty affects
  the $\Upsilon(1S)$ and $\Upsilon(2S)+\Upsilon(3S)$ measurements in a fully correlated way.
  The lower limits for $R(\chi_{b})$ and $\lambda_{\vartheta}(\Upsilon\mathrm{\:from\:}\chi_{b})$
  depend on the ratio $R(\chi_{b2})/R(\chi_{b1})$, for which three
  different values are assumed. }
  \label{fig:E866_feeddown}
\end{figure}
We will base our second scenario, for the bottomonium family, on the precise
and detailed measurement of E866\cite{bib:e866_Ups}, shown in
Fig.~\ref{fig:E866_feeddown}a. This result offers several interesting cues. It
is remarkable that the $\Upsilon(2S)$ and $\Upsilon(3S)$ are found to be almost
fully polarized, while the $\Upsilon(1S)$ is only weakly polarized. The most
reasonable explanation of this fact is that the fraction of $\Upsilon(1S)$
mesons coming from $\chi_{b}$ decays is large and its polarization is very
different with respect to the polarization of the directly produced
$\Upsilon(1S)$. In fact, in the assumption that all directly produced $S$
states have the same polarization, we can translate the E866 measurement into a
lower limit for the feed-down fraction $R(\chi_{b})$ from $P$ states, summing
together $1P_1$, $1P_2$, $2P_1$, $2P_2$ contributions. We assume that the
$\Upsilon(2S)+\Upsilon(3S)$ result has a negligible contamination from
$\chi_{b}^\prime \rightarrow \Upsilon(2S) \gamma$ decays and, therefore,
provides a good evaluation of the polarization of the directly produced $S$
states (a conservative assumption for this specific calculation). The lower
limit for $R(\chi_{b})$ corresponds to the case $J_z(\chi_{b1}) =
J_z(\chi_{b1}^\prime) = \pm 1$, $J_z(\chi_{b2}) = J_z(\chi_{b2}^\prime) = 0$,
in which the $\Upsilon(1S)$ mesons from $\chi_{b}$ decays have the largest
negative value of $\lambda_{\vartheta}$. The result, depending slightly on the
assumed ratio between $P_2$ and $P_1$ feed-down contributions, is shown in
Fig.~\ref{fig:E866_feeddown}b as a function of $p_{\rm T}$. More than $50\%$ of
the $\Upsilon(1S)$ are produced from $P$ states for $\langle p_{\mathrm{T}}
\rangle \simeq 0.5$~GeV$/c$, and more than $30\%$ for $\langle p_{\mathrm{T}}
\rangle \simeq 2.3$~GeV$/c$. These limits are appreciably higher than the value
of the feed-down fraction of J/$\psi$ from $\chi_{c}$ measured at similar
energy, low $p_{\rm T}$ and mid rapidity\cite{bib:HERAB_chic}. We remind that
we have obtained only a lower limit (no upper limit is implied by the data),
corresponding to the case in which $\chi_{b1}$ and $\chi_{b2}$ are always
produced in the same very specific and pure angular momentum configurations.
Any deviation from this extreme case would lead to higher values of the
indirectly determined feed-down fraction.

The E866 measurement data also set an upper limit on the combined polarization
of $\chi_{b1}$ and $\chi_{b2}$. Figure~\ref{fig:E866_feeddown}c shows the
derived range of possible polarizations of $\Upsilon(1S)$ coming from
$\chi_{b}$. The upper bound, corresponding to $R(\chi_{b}) = 1$, coincides with
the measured $\Upsilon(1S)$ polarization. The lower bound, slightly depending
on the relative contribution of $\chi_{b1}$ and $\chi_{b2}$, is not influenced
by the E866 data and corresponds to the minimum ($p_{\mathrm{T}}$ dependent)
value of $R(\chi_{b})$ represented in Fig.~\ref{fig:E866_feeddown}b. The second
strong indication of the E866 data is, therefore, that at low $p_{\rm T}$ the
$\Upsilon(1S)$ coming from $\chi_{b}$ decays has a longitudinal component in
the CS frame larger than $\sim 30\%$ ($\lambda_{\vartheta} \lesssim 0.1$),
being $\sim 60\%$ ($\lambda_{\vartheta} \sim -0.5$) the maximum amount of
longitudinal polarization that the $\Upsilon(1S)$ produced in this way is
allowed to have.

In the light of these scenarios it is clear that measurements of the
polarization of the $\chi$ states will be extremely important for an
unambiguous understanding of the J/$\psi$ and $\Upsilon(1S)$
polarizations.

\begin{figure}[htb]
\centering
\includegraphics[width=0.4\linewidth]{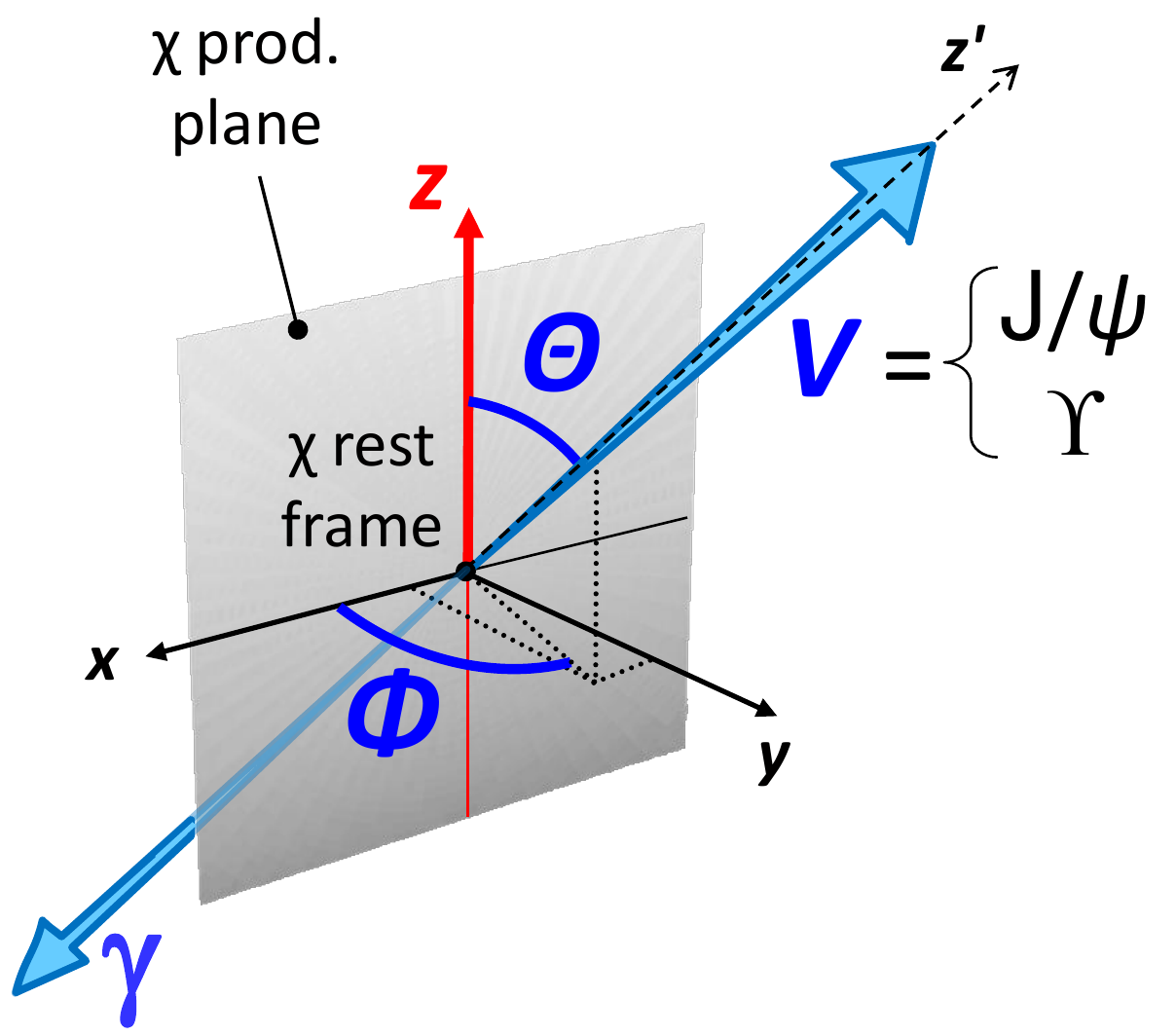}
\caption{Definition of axes and decay angles for $\chi \rightarrow V
\gamma$.} \label{fig:chinotation}
\end{figure}

\begin{figure}[htb]
\centering
\includegraphics[width=0.9\linewidth]{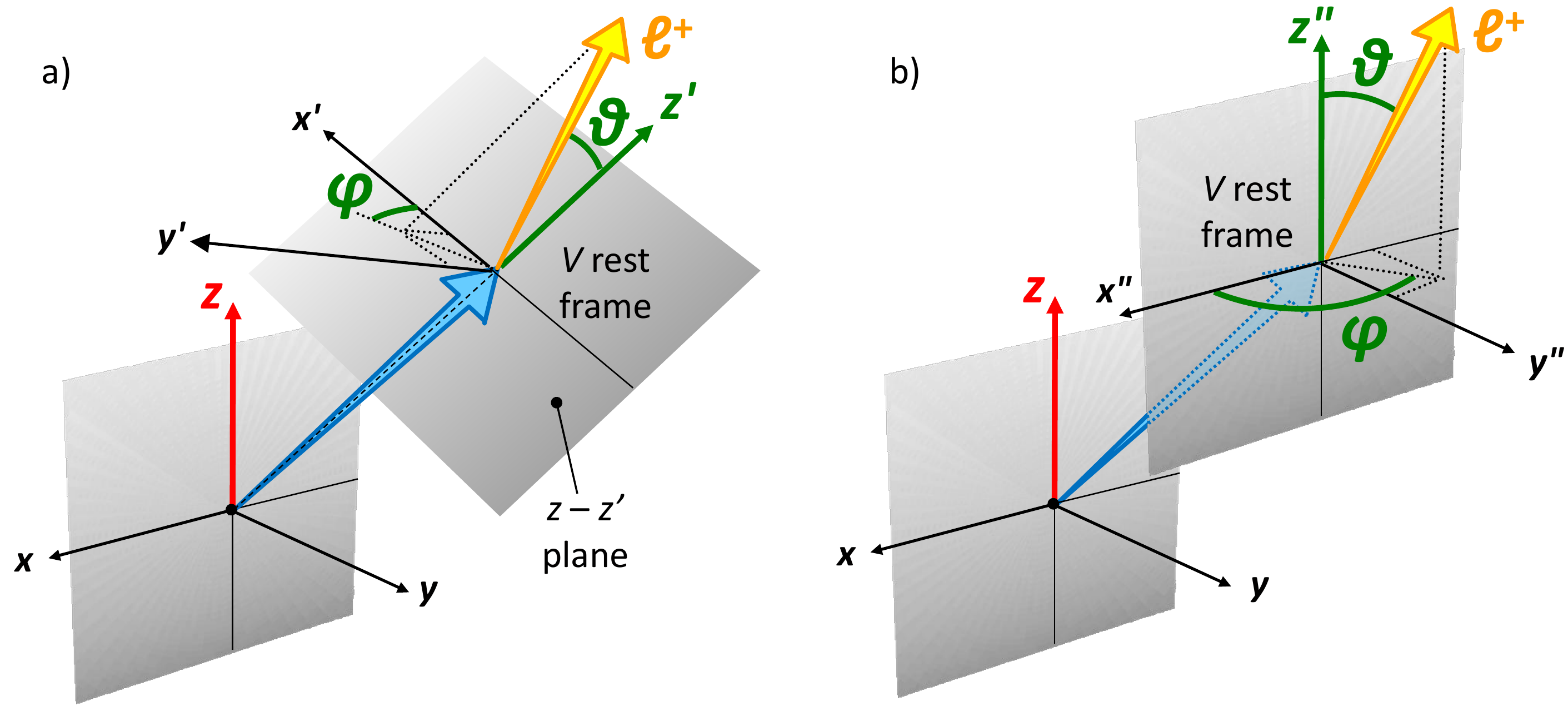}
\caption{The $V \rightarrow \ell^+ \ell^-$ decay angles with two definitions of
the $V$ polarization axis: parallel to the $V$ momentum direction in the $\chi$
rest frame (a), or parallel to the $\chi$ polarization axis (b).}
\label{fig:chiframes}
\end{figure}

It has been shown in Ref.~\refcite{bib:chi_polarization} that the angular
momentum compositions of the \chiAll\ states produced in high energy collisions
can be derived from the dilepton decay distributions of the daughter \jpsi\ or
\upsAll\ mesons, with a reduced dependence on the details of reconstruction and
simulation of the radiated photon. This method is based on a particular choice
of the quantization axes. Different frame definitions are in principle suitable
for $\chi_{c}$ and $\chi_{b}$ polarization studies in hadronic collisions. We
generically denote by $V$ the charmonium and bottomonium $^3\!S_1$ states,
$\mathrm{J}/\psi$ and $\Upsilon$, and by $\chi$ the $^3\!P_{j}$ states,
$\chi_{cj}$ and $\chi_{bj}$, with $j=1,2$. Notations for axes and angles for
the description of the $\chi \rightarrow V \gamma$ decay are defined in
Fig.~\ref{fig:chinotation}, where $z$ is the $\chi$ polarization axis (for
example the HX or CS axes, defined in the $\chi$ rest frame).
The traditional choice of axes, adopted in calculations and
measurements (cited in Ref.~\refcite{bib:chi_polarization}) of the
full decay angular distribution for \chicAll\ mesons produced at low
laboratory momentum, is represented in Fig.~\ref{fig:chiframes}(a),
where the $V$ polarization axis, $z^\prime$, is the $V$ direction in
the \chiAll\ rest frame. With respect to this system of axes, taking
the polar anisotropy parameter $\lambda_\vartheta$ as an example,
all measurements will find, for \chiOne\ and \chiTwo\ dileptons, the
values
\begin{equation}
\lambda_\vartheta^{j=1} = -\frac{1}{3} \left[ 1 - \frac{16}{3} h_2 +
\mathcal{O}(h_2^2) \right] \quad {\rm and} \quad \lambda_\vartheta^{j=2} =
\frac{1}{13} \left[ 1 - \frac{80 \sqrt{5}}{13} g_2 + \mathcal{O}(g_2^2) \right]
\, ,
\end{equation}
where $h_2$ and $g_2$ are the fractional contributions of the
magnetic quadrupole transitions (electric octupole transitions for
the $j=2$ case have been neglected). The dilepton distribution in
the $x^{\prime},y^{\prime},z^{\prime}$ coordinate system is
\emph{independent} of the \chiAll\ polarization state. This choice
of axes is suitable for measuring the contribution of the
higher-order multipoles, but it does not provide information on the
polarization of the \chiAll\ and, hence, on its production
mechanism, when the photon distribution is integrated out.
An alternative definition, proposed in
Ref.~\refcite{bib:chi_polarization}, enables the determination of
the \chiAll\ polarization in high-momentum experiments without the
need of measuring the full photon-dilepton kinematic correlations.
This definition, shown in Fig.~\ref{fig:chiframes}(b), ``clones''
the \chiAll\ polarization frame, defined in the \chiAll\ rest frame,
into the $V$ rest frame, taking the
$x^{\prime\prime},y^{\prime\prime},z^{\prime\prime}$ axes to be
parallel to the $x,y,z$ axes. As explained in
Ref.~\refcite{bib:chi_polarization}, the dilepton distribution in
this frame contains as much information as the photon distribution
regarding the \chiAll\ polarization state: the two distributions are
even \emph{identical} when higher-order multipoles are neglected.

The definition of the $x,y,z$ axes (and, therefore, of the
$x^{\prime\prime},y^{\prime\prime},z^{\prime\prime}$ axes) uses the momenta of
the colliding hadrons \emph{as seen in the \chiAll\ rest frame}, so that it
requires, in general, the knowledge of the photon momentum. However, \emph{for
sufficiently high (total) momentum of the dilepton}, the \chiAll\ and $V$ rest
frames coincide and the $x^{\prime\prime},y^{\prime\prime},z^{\prime\prime}$
axes can be approximately defined using only momenta seen in the $V$ rest
frame. For example, if the \chiAll\ polarization axis ($z$) is defined along
the bisector of the beam momenta in the \chiAll\ rest frame (CS frame), the
corresponding $z^{\prime\prime}$ axis is approximated by the bisector of the
beam momenta in the $V$ rest frame.
The relative error induced by this approximation on the polar anisotropy
parameter is $\left| \Delta\lambda_\vartheta / \lambda_\vartheta \right| =
\mathcal{O}\left[ (\Delta M / p )^2 \right]$,
where $\Delta M$ is the $\chi - V$ mass difference and $p$ is the \emph{total}
laboratory momentum of the dilepton. Therefore, for not-too-small momentum this
frame definition coincides with the frame defined in the measurement of the
polarization of inclusively produced \jpsi~/~\upsAll\ mesons (CS or HX, for
example). In other words, the measurement of the dilepton distribution at
sufficiently high laboratory momentum provides a direct determination of the
\chiAll\ polarization along the chosen polarization axis. This determination is
cleaner than the one using the photon distribution in the \chiAll\ rest frame,
because it is independent of the knowledge of the higher-order photon
multipoles.

We remark that the general equations describing the \chiAll\ angular
decay distributions \emph{depend} on the exact definition of the
quantization axes of \chiAll\ and $V$. An incorrect match between
equations and frame definitions created confusion in the past,
probably leading to wrong measurements\cite{bib:chi_polarization}.
While reading the recent LHCb paper\cite{bib:LHCb} of $\chi_{c}$
production in pp collisions we wonder if we might be in the presence
of a similar situation. To calculate the effect of the unknown
$\chi_{c1}$ and $\chi_{c2}$ polarizations on the calculation of the
$\chi_{c}$ and $\mathrm{J}/\psi$ reconstruction and selection
efficiencies, the authors reweight the simulated events assuming
different $\chi_{c1}$ and $\chi_{c2}$ polarization scenarios. The
formulas used for the angular distributions are those listed in the
HERA-B paper on $\chi_{c}$ production\cite{bib:HERAB_chic}, where
the choice of the polarization axes is of the type represented in
Fig.~\ref{fig:chiframes}(a). However, the axis definitions used in
the LHCb paper seem to correspond to the convention represented in
Fig.~\ref{fig:chiframes}(b), considered in the high-momentum limit
(certainly an excellent approximation in the LHCb case). First they
define, to describe the prompt-$\mathrm{J}/\psi$ decays, the angle
$\theta_{\mathrm{J}/\psi}$ as ``the angle between the directions of
the $\mu^+$ in the $\mathrm{J}/\psi$ rest frame and [of] the
$\mathrm{J}/\psi$ in the laboratory frame'': that is, the chosen
$\mathrm{J}/\psi$ polarization axis is the centre-of-mass helicity
frame. What perplexes us is the subsequent definition of the angles
of the $\chi_{c} \rightarrow \mathrm{J}/\psi \gamma \rightarrow
\ell^+ \ell^- \gamma$ decay chain: ``The $\chi_{c} \rightarrow
\mathrm{J}/\psi \gamma$ system is described by
$\theta_{\mathrm{J}/\psi}$ and two further angles, $\theta_{\chi_c}$
and $\phi$, where $\theta_{\chi_c}$ is the angle between the
directions of the $\mathrm{J}/\psi$ in the $\chi_c$ rest frame and
[of] the $\chi_c$ in the laboratory frame [i.e., the angle $\Theta$
of Fig.~\ref{fig:chinotation}]''. This means that
$\theta_{\mathrm{J}/\psi}$, referred to the $\mathrm{J}/\psi$
helicity axis in the laboratory, is taken as polar angle of the
dilepton decays, while, to be consistent with the formulas used, a
new angle $\theta_{\mathrm{J}/\psi}^\prime$ should be introduced and
defined as the angle between the directions of the $\mu^+$ in the
$\mathrm{J}/\psi$ rest frame and of the $\mathrm{J}/\psi$ in the
$\chi_c$ rest frame. The dilepton azimuthal angle $\phi$, now
defined as ``the angle between the plane formed (\dots) [by] the
$\chi_c$ and $\mathrm{J}/\psi$ momentum vectors in the laboratory
frame and the $\mathrm{J}/\psi$ decay plane in the $\mathrm{J}/\psi$
rest frame'', should rather be defined as ``the angle between the
$\mathrm{J}/\psi$ decay plane \emph{in the $\chi_c$ rest frame} [the
two leptons are collinear in the $\mathrm{J}/\psi$ rest frame] and
the plane formed by the $\chi_c$ direction in the laboratory frame
[being the helicity axis the chosen quantization axis for the
$\chi_c$] and the $\mathrm{J}/\psi$ direction in the $\chi_c$ rest
frame''.

We emphasize that such an inconsistency between axis definitions and formulas
results in a completely wrong description of the angular distributions. The
sentence ``The angular distributions are independent of the choice of
polarisation axis'' may indicate a crucial misunderstanding: it is true that
the angular distributions do not depend in form on the choice of the $\chi_c$
quantization axis, but they depend drastically, also in form, on the choice of
the $\mathrm{J}/\psi$ quantization axis. Let us consider, for example, the
dilepton distribution, integrated over the photon distribution, in the
$\chi_{c2}$ case. With the choice of axes made in the LHCb paper, and using the
correct formulas, the cases of $\chi_{c2}$ having helicity $\pm2$ or $0$ would
be observed, respectively, as a fully transverse ($\lambda_\vartheta = +1$) or
dominantly longitudinal ($\lambda_\vartheta = -3/5$) $\mathrm{J}/\psi$
polarization. Instead, the inconsistent formulas predict (erroneously, given
the mismatch with the definition of the axes) an almost isotropic decay
distribution ($\lambda_\vartheta = +1/13$), independently of the $\chi_{c2}$
helicity, as previously discussed.

\section{Polarization as an indication of sequential
suppression} \label{sec:seqsuppr}

Hypotheses on the suppression of $\chi_c$ and $\chi_b$ production in
nucleus-nucleus collisions play a crucial role in the interpretation of the
J/$\psi$ and $\Upsilon(1S)$ measurements from
SPS\cite{bib:NA38,bib:NA50,bib:NA60},
RHIC\cite{bib:PHENIX1,bib:PHENIX2,bib:PHENIX3,bib:STAR} and
LHC\cite{bib:ATLAS,bib:CMS,bib:ALICE} in terms of evidence of quark-gluon
plasma (QGP) formation. The observation of the $\chi_c$ and $\chi_b$
suppression patterns in Pb-Pb collisions at the LHC could confirm or falsify
the ``sequential quarkonium melting'' scenario\cite{bib:sqm,bib:Karsch} and,
therefore, discriminate between the QGP interpretation and other options.
However, a direct observation of the $\chi_c$ and $\chi_b$ signals in their
radiative decays to ${\rm J}/\psi$ and $\Upsilon(1S)$ is practically impossible
in heavy-ion collisions, given the very large number of background photons
produced in such events.

\begin{figure}[tbh]
  \centering
  \includegraphics[width=0.495\textwidth]{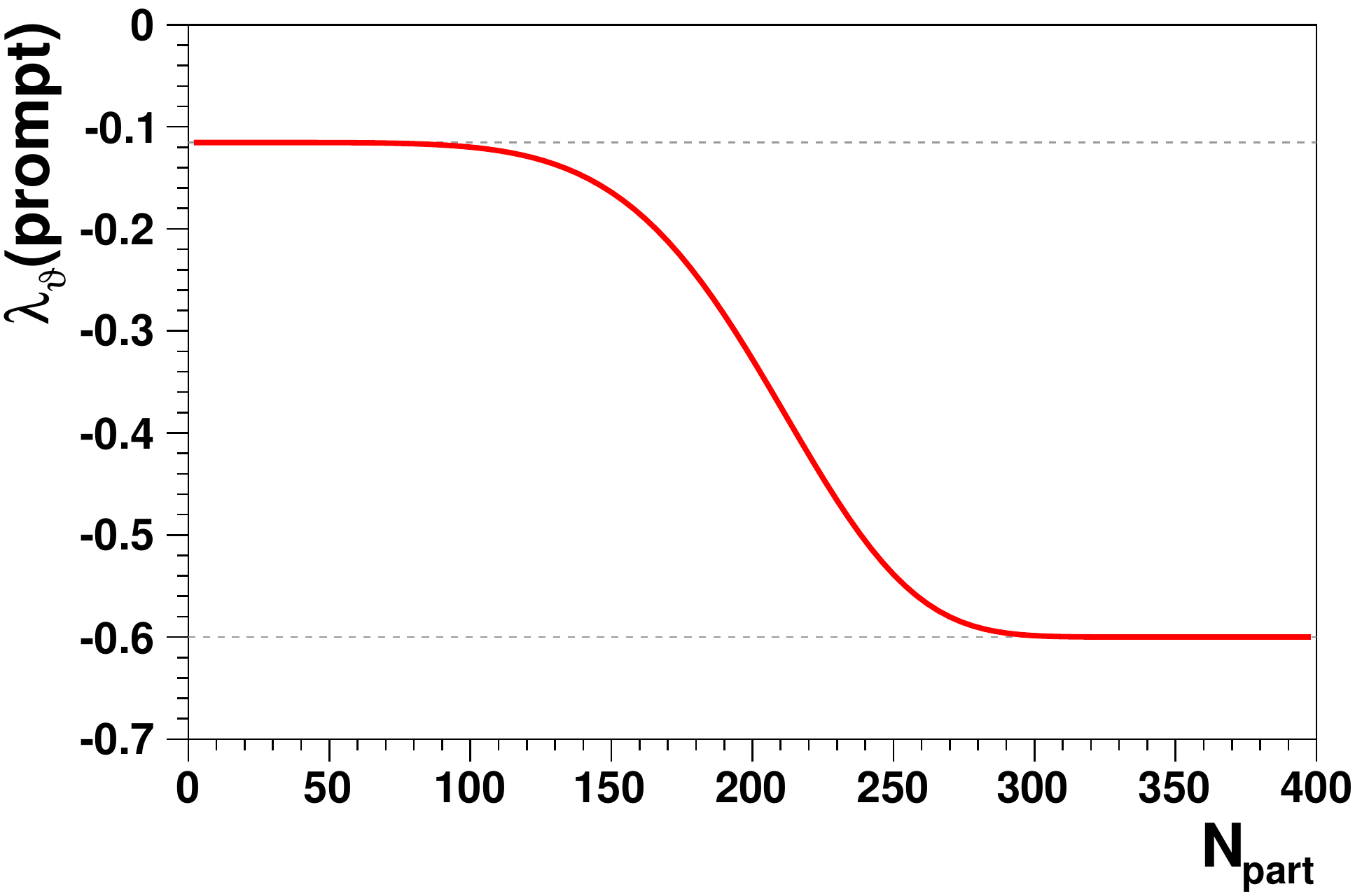}
  \includegraphics[width=0.495\textwidth]{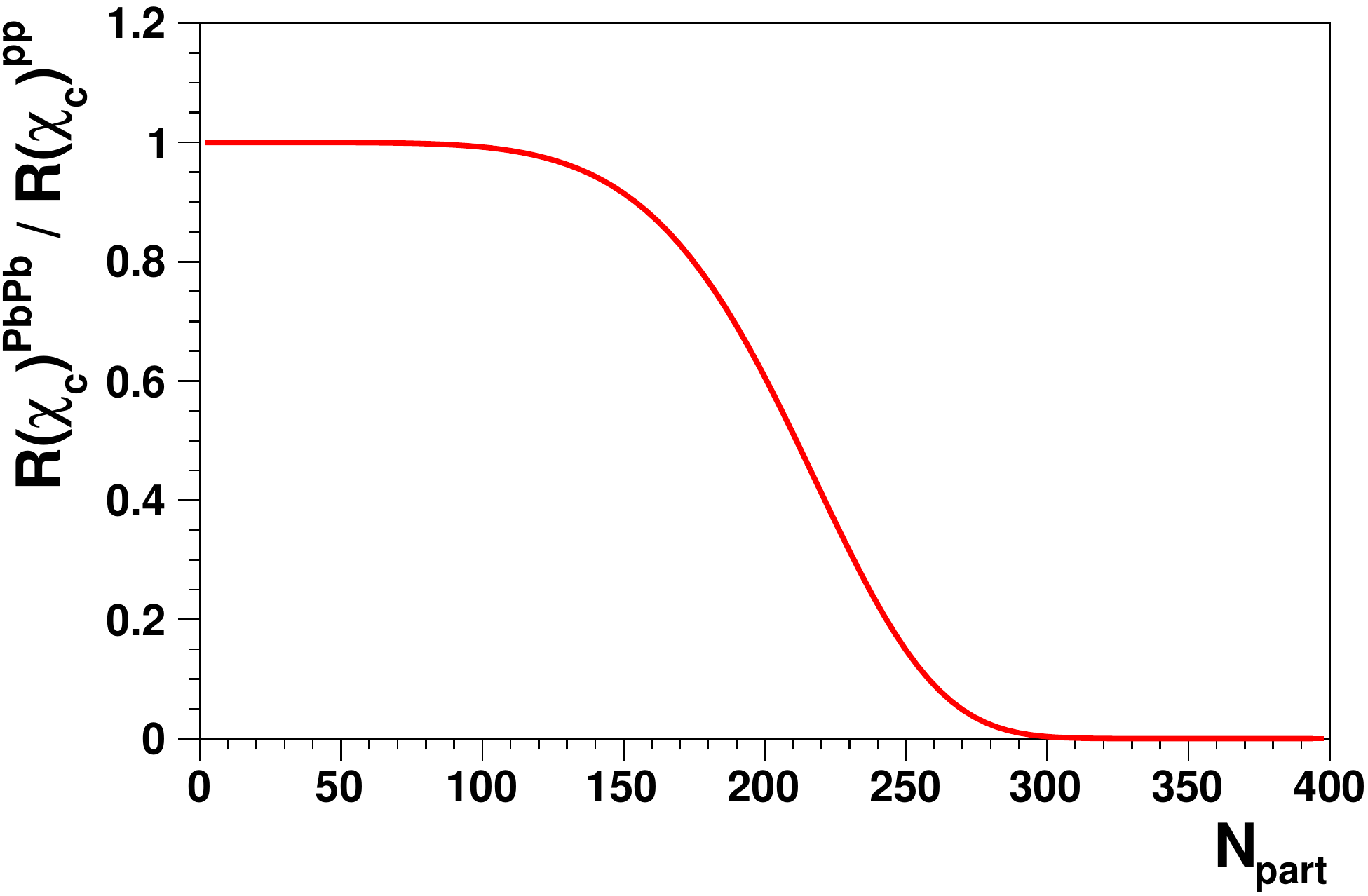}
  \caption{ A hypothetical variation of $R(\chi_c)$ (normalized to the pp value)
  with the centrality of the Pb-Pb collision (left) and the consequent variation
  of the prompt-J/$\psi$ polarization $\lambda_\vartheta$
  (right), according to the charmonium polarization scenario discussed in the text. }
  \label{fig:pol_seq_suppr}
\end{figure}

The E866 scenario suggests an alternative method to determine the relative
yield of \emph{P} and \emph{S} states by performing only dilepton polarization
measurements. This possibility is particularly valuable in the perspective of
quarkonium measurements in heavy-ion collisions. A change of the observed
J/$\psi$ and $\Upsilon(1S)$ polarizations from proton-proton to central
nucleus-nucleus collisions would directly reflect differences in the nuclear
dissociation patterns of $S$ and $P$ states.\cite{bib:chiPolSuppr}
Figure~\ref{fig:pol_seq_suppr} illustrates the concept of the method. The left
panel shows an hypothetical $R(\chi_c)$ pattern inspired from the sequential
charmonium suppression scenario, in which the $\chi_c$ yield disappears rapidly
beyond a critical value of the number of nucleons participating in the
interaction ($N_{\rm part}$). This effect would be reflected by a change in the
observed prompt-J/$\psi$ polarization. As shown in the right panel, according
to the scenario presented in Fig.~\ref{fig:CDF_direct_extrapolation} the
polarization should become significantly more longitudinal (in the helicity
frame) after the disappearance of the transversely polarized feed-down
contribution due to $\chi_c$ decays. We are assuming that the ``base''
polarizations of the directly produced $S$ and $P$ states remain essentially
unaffected by the nuclear medium and are, therefore, not distinguishable from
those measurable in pp collisions. A test of the sequential suppression pattern
can, therefore, be made by comparing the prompt-J/$\psi$ polarization measured
in pp (or peripheral nucleus-nucleus) collisions with the one measured in
central nucleus-nucleus collisions and checking that this latter tends to the
polarization of the \emph{directly} produced $S$ states, also determined in pp
collisions through $\psi^\prime$ measurements.

The same method can be applied to the measurement of $\chi_b$ suppression using
$\Upsilon(1S)$ polarization. According to the E866 scenario
(Fig.~\ref{fig:E866_feeddown}), in pp (and peripheral Pb-Pb) collisions the
$\Upsilon(1S)$ should be only slightly polarized, reflecting the mixture of
directly and indirectly produced states with opposite polarizations. In central
Pb-Pb collisions the $\Upsilon(1S)$ would acquire the fully transverse
polarization characteristic of the directly produced $S$ states, indicating the
suppression of the $P$ states.

We have estimated that about $30$k prompt-J/$\psi$ and $10$k $\Upsilon(1S)$
signal events, with both leptons having $p_{\rm T}
> 5$~GeV$/c$ and an assumed background fraction of $40\%$, would lead to a
significant indication of the nuclear disassociation of the $\chi$ states
according to the scenarios we have considered.

\section{Summary}
\label{sec:summ}

Several puzzles affect the existing measurements of quarkonium polarization.
The experimental determination of the \jpsi\ and \upsAll\ polarizations must be
improved.

Measurements and calculations of vector quarkonium polarization
should provide results for the full dilepton decay angular
distribution (a three-parameter function) and not only for the polar
anisotropy parameter. Only in this way can the measurements and
calculations represent unambiguous determinations of the average
angular momentum composition of the produced quarkonium state in
terms of the three base eigenstates, with $J_z = +1, 0, -1$.

Moreover, it is advisable to perform the experimental analyses in at
least two different polarization frames. In fact, the self-evidence
of certain signature polarization cases (e.g.\ a full polarization
with respect to a specific axis) can be spoiled by an unfortunate
choice of the reference frame, which can lead to artificial
(``extrinsic'') dependencies of the results on the kinematics and on
the experimental acceptance.

The angular distribution can be characterized by a frame-independent
quantity, $\tilde{\lambda}$, calculable in terms of the polar and
azimuthal anisotropy parameters.  This frame-invariant observable
can be used during the data analysis phase to perform
self-consistency checks that can expose previously unaccounted
biases, caused, for instance, by the detector limitations or by the
event selection criteria. The variable $\tilde{\lambda}$ also
provides relevant physical information: it characterizes the
\emph{shape} of the angular distribution, reflecting ``intrinsic''
spin-alignment properties of the decaying state, irrespectively of
the specific geometrical framework chosen by the observer. Extrinsic
dependencies on kinematics and acceptances are cancelled exactly,
enabling more robust comparisons with other experiments and with
theory.

Stripped-down analyses which only measure the polar anisotropy in a
single reference frame, as often done in past experiments, give more
information about the frame selected by the analyst (``is the
adopted quantization direction an optimal choice?'') than about the
physical properties of the produced quarkonium (``along which
direction is the spin aligned, on average?'').
For example, a natural longitudinal polarization will give any desired
$\lambda_\vartheta$ value, from $-1$ to $+1$, if observed from a suitably
chosen reference frame.  Lack of statistics is not a reason to ``reduce the
number of free parameters'' if the resulting measurements become ambiguous.

Besides improving methodology aspects, more detailed and elementary information
will have to be provided, by measuring separately the polarizations of directly
and indirectly produced states.

The forthcoming measurements of quarkonium polarization in proton-proton
collisions at the LHC have the potential of providing a very important step
forward in our understanding of quarkonium production, if the experiments adopt
a more robust analysis framework, incorporating the ideas presented here.

Quarkonium polarization can also be used as a new probe for the formation of a
deconfined medium. This method, based on the study of dilepton kinematics
alone, provides a feasible and clean alternative to the direct measurement of
the $\chi$ yields through reconstruction of radiative decays. With sizeable
J/$\psi$ and $\Upsilon(1S)$ event samples to be collected in nucleus-nucleus
collisions, the LHC experiments have the potential to provide a clear insight
into the role of the $\chi$ states in the dissociation of quarkonia, a crucial
step forward in establishing the validity of the sequential melting mechanism.

\section*{Acknowledgments}

It is a pleasure to acknowledge a very fruitful collaboration with my
colleagues and friends C. Louren\c{c}o, J. Seixas and H. W\"{o}hri. I thank the
support of Funda\c{c}\~ao para a Ci\^encia e a Tecnologia, Portugal (contracts
SFRH/BPD/42343/2007, CERN/FP/116367/2010, CERN/FP/116379/2010).

\end{document}